\documentstyle[emulateapj,psfig,epsf]{article}
\def\kms{km\,s$^{-1}$\,}
\def\vs{$V_{S}$\,\,}
\def\etal{ et~al.\rm\,}

\def\snr{0505$-$67.9\,}
\def\s2{DEM L 71\,}
\def\chan{{\it Chandra\,}}
\def\bn{$I_{B}/I_{N}$\,}
\def\tetp{($T_{e}/T_{p}$)$_{0}$\,}
\def\fneut{$f_{H^{0}}$\,}
\def\memp{$m_{e}/m_{p}$\,}

\tighten
\begin{document}

\title{The Physics of Supernova Blast Waves. I. Kinematics of
\s2 in the Large Magellanic Cloud}

\author{Parviz Ghavamian, Cara E. Rakowski,
John P. Hughes\altaffilmark{1} and T. B. Williams\altaffilmark{1} }

\affil{Department of Physics and Astronomy, Rutgers University, Piscataway, NJ 08854-8019;
parviz,rakowski,jph,williams@physics.rutgers.edu}

\submitted{Accepted March 4, 2003}

\begin{abstract}

We present the results from Fabry-Perot imaging spectroscopy of the Balmer-dominated supernova remnant
\s2 (\snr) in the LMC.  Spectra extracted from the
entire circumference of the blast wave reveal the broad and narrow component H$\alpha$ line
emission characteristic of non-radiative shocks in partially neutral gas.  The new spectra of \s2 include
portions of the rim that have not been previously observed.  We find that the broad component width varies
azimuthally along the edge of \s2, ranging from 450$\pm$60 \kms along the eastern edge
to values as high as 985$^{+210}_{-165}$ \kms along the faint western edge.  The latter width is nearly
60\% larger than the value determined by earlier spectroscopy of the brightest Balmer-dominated filaments.  In part of the
faint northern rim the broad component is not detected, possibly indicating a lower density in these regions and/or 
a broad component width in excess of 1000 \kms.  Between the limits of zero and full electron-ion temperature equilibration at
the shock front, the allowed range of shock velocities is 430$-$560 \kms along
the east rim and 700$-$1250 \kms along other parts of the blast wave.
The H$\alpha$ broad-to-narrow flux ratios vary considerably around the remnant, ranging from
0.4 to 0.8.  These ratios lie below the values predicted by our shock models.  We find that narrow component H$\alpha$
emission from a cosmic ray precursor may be the cause of the discrepancy.  The least decelerated portions of
the blast wave (i.e., regions excluding the brightest filaments) are well characterized by Sedov models with a kinetic
energy $E_{51}$\,=\,(0.37$\pm$0.06)\,$D^{\,5/2}_{\,50}$, where $D_{\,50}$ is the LMC distance in units of 50 kpc.  
The corresponding age for \s2 is (4360$\pm$290)\,$D_{\,50}$ yr.  This is the first time that velocity information
from the entire blast wave has been utilized to study the global kinematics of a non-radiative SNR at a known distance.

\keywords{ ISM: supernova remnants: individual (\snr, \s2)--ISM: kinematics and dynamics, shock waves}
\end{abstract}
 
\altaffiltext{1}{Visiting Astronomer, Cerro Tololo Inter-American Observatory, National Optical Astronomy
Observatories.  CTIO is operated by AURA, Inc.\ under contract to the National Science
Foundation.}

\section{INTRODUCTION}

One distinctive property of non-radiative supernova remnants (SNRs) in partially neutral 
gas is the presence of optical line emission which allows us to trace the precise
positions of their blast waves.  The optical spectra of these SNRs (termed non-radiative because
postshock cooling losses are negligible) are dominated by Balmer line
emission, produced by collisional excitation when neutral hydrogen is overrun by the blast
wave (Chevalier \& Raymond 1978, Bychkov \& Lebedev 1979).  Each emission line consists of two components:
(1) a narrow velocity component produced when cold, ambient H~I overrun by the shock is excited by electron and proton collisions,
and (2) a broad velocity component produced when fast neutrals created by postshock charge exchange are
collisionally excited (Chevalier, Kirshner \& Raymond 1980).  Since the optical emission arises in a very thin
($\lesssim\,$10$^{16}$ cm) ionization zone, measuring the width of the broad Balmer
line yields the proton temperature immediately behind the shock.  The broad-to-narrow flux
ratio, on the other hand, is sensitive to both the initial electron-ion temperature equilibration at the shock front
and the neutral fraction of the preshock gas (Chevalier, Kirshner \& Raymond 1980,
Smith \etal\, 1991, hereafter SKBW91). Together, the broad-to-narrow
ratio and broad component width are highly useful tools for estimating the shock velocity,
degree of electron-ion equilibration and even the preshock neutral fraction in
SNRs (SKBW91, Ghavamian 1999, Ghavamian \etal\, 2001; hereafter GRSH01, Ghavamian \etal\, 2002; hereafter GRHB00).

One of four known Balmer-dominated SNRs in the LMC, \s2 was originally discovered as an
unresolved X-ray source by {\it Einstein} (Long, Helfand
\& Grabelsky 1981).  The remnant appears as a limb brightened shell, tear-drop shaped and
approximately 1.4\arcmin\,$\times$\,1.2\arcmin\, in size.
Since its discovery, \s2 has been studied in the optical via narrow band imagery (Tuohy \etal\,
1982) and low and high resolution longslit spectroscopy (SKBW91, Smith \etal\, 1994, hereafter SRL94).  At X-ray energies,
the remnant has been studied spectroscopically with {\it ASCA} SIS (Hughes, Hayashi \& Koyama 1998; hereafter HHK98) 
and {\it Chandra} ACIS-S (Hughes \etal\, 2002, hereafter HGRS03; Rakowski \etal\, 2002, hereafter RGH03).  In their
analysis of the global X-ray spectrum, HHK98 assumed a Sedov dynamical evolution and
applied self-consistent non-equilibrium ionization models to derive physical parameters for
\s2 and six other remnants in the LMC.  They found that the Fe abundance in \s2
was twice as high as that of the other remnants in the sample, leading them to conclude that \s2
is the remnant of a Type Ia explosion.  In a more recent {\it Chandra} analysis of spatially resolved 
ejecta material, HGRS03 again measured a highly elevated abundance of Fe and other Si-group
elements, making an even stronger case for a Type Ia origin of \s2.

While X-ray analyses have allowed us to build a global picture of \s2, optical spectroscopy
has thus far been limited to longslit observations passing through the bright east/west limbs of the SNR (see
Figure~1).  In these cases the slit was oriented east/west along the diameter of the remnant, resulting
in detections of broad ($\sim\,$600 \kms FWHM) and narrow H$\alpha$ emission from the brightest optical filaments and
some of the fainter emission between them (SKBW91, SRL94).  However, 
both Tuohy \etal\, (1982) and SKBW91 noted that the broad and narrow H$\alpha$ emission from
the bright filaments was accompanied by moderate [S~II], [N~II] emission, a sign that
the gas behind these shocks has begun to cool.  In this case, since the narrow component H$\alpha$ emission 
is produced both by collisional excitation at the shock front {\it and} recombination of the cooling gas,
the broad-to-narrow ratios of the bright filaments cannot be used to infer the shock parameters.

Not all of the bright H$\alpha$ in \s2 is confined to filaments.
Prominent, localized lumps of emission can also be seen in the interior.  Three particularly bright patches
around 5\arcsec in size are located 10\arcsec$-$20\arcsec\, inside the west rim. (Figure~1).  These patches are also visible in 
narrow band [S~II] images of \s2.  At least half a dozen other lumps can also be seen scattered throughout the interior of \s2.
Interestingly, a comparison of our narrowband H$\alpha$ images with {\it Chandra} images of \s2 reveals that the brightest of
the interior patches have no X-ray counterparts (RGH03).   The optical emission from these clouds may be
produced by clumpy neutral gas which has  been photoionized by He~II $\lambda$304 emission from
the blast wave (GRHB00).

Given the difficulty of separating radiative from non-radiative shocks in prior optical spectra
of \s2, we undertook a new spectroscopic study to investigate the more widely distributed
optical emission around this remnant.  While considerably fainter than the prominent E/W
filaments, the fainter emission extends unbroken over a larger portion of the circumference and
traces pure Balmer-dominated shocks, as evidenced by the lack of [S~II] and [O~III] emission
from these shocks in the images of Tuohy \etal\, (1982) (and our own unpublished imagery). 
Assuming that variations in H$\alpha$ surface brightness reflect density variations in the surrounding ISM,
the fainter emission should trace portions of the blast wave that have suffered the least deceleration.
To investigate the entire network of Balmer-dominated shocks in \s2, we observed the remnant with the
Rutgers Fabry-Perot imaging spectrometer (RFP), centered on the narrow component H$\alpha$ line at the velocity
of the LMC.  Since the entire SNR (1.4\arcmin\,$\times$\,1.2\arcmin) lay within the RFP field of view,
we were able to extract spectra from the entire blast wave rim while simultaneously
measuring the radius of each extraction point from the center of \s2. This was invaluable for 
correlating the variation of shock speed with both radius and position angle along the
blast wave rim.

Combined with the recent observation of \s2 by {\it Chandra} (HGRS03, RGH03)
the RFP spectra present us with an unprecedented opportunity to probe the underlying physics
of high Mach number, collisionless shocks.  Thanks to its 0.5\arcsec\, resolution, {\it Chandra} 
has revealed a strikingly detailed correlation between the X-ray and optical morphologies of the blast
wave, suggesting a common origin for both sources of emission (HGRS03, RGH03).  As shown by RGH03,
the postshock proton temperatures and shock velocities measured from Balmer-dominated spectra can be combined
with the postshock electron temperatures of the blast wave measured from {\it Chandra} spectra to measure the
equilibration fraction immediately behind the shock (\tetp) as a function of shock speed.  As we will show in this
paper, the RFP and {\it Chandra} data probe shock speeds that vary by a factor $\sim$ 3 along the rim of 
\s2, therefore we can use our multiwavelength study of this SNR to probe the (poorly known) degree of 
collisionless heating over a large sample of shock speeds, in a model-independent manner.

\section{SPECTROSCOPIC OBSERVATIONS}

Our observations of \s2 were performed with the RFP spectrometer at the 1.5 m telescope of Cerro Tololo 
Inter-American Observatory on 1998 February 18 and February 21 (UT).  The Tek1024 CCD and 200 mm camera
lens were used with the f/7.5 secondary focus, giving an image scale of 0\farcs65
pixel$^{-1}$. The field of view was a circle 7.8\arcmin\, in diameter, centered on coordinates
05$^{\rm h}$05$^{\rm m}$42\fs8, $-$67$^{\circ}$52\arcmin35\farcs9 (2000).  
The scans were centered on the LMC systemic velocity $V_{HELIO}(LMC)$\,=\,+278 \kms and sampled the
H$\alpha$ line profile at 14 velocity slices $\Delta$ $V_{i}$\,= ($-$704, $-$617, $-$480, $-$347,
$-$210, $-$155, $-$128, $-$77, $-$18, +87, +192, +330, +471, +791) \kms.  All scans 
except for those at $-$704 and +791 \kms were performed on the first night of
the observing run.  One image was acquired at each etalon setting, with an exposure time of
750s at $-$704 \kms and +791 \kms and 500s at all other settings.  The average seeing 
during the first night was 2\farcs2 FWHM; the
worst seeing occurred during the scans centered redward of the H$\alpha$ narrow component.
During the second night, the average seeing was 2\farcs8 FWHM.  In our observations
we used the CTIO H$\alpha$ filter 6563/75 ($\lambda_{FWHM}$\,=\,75 \AA) to isolate emission from
a single spectral order.

We reduced the Fabry-Perot images using IRAF\footnote{IRAF is distributed by the National Optical Astronomy Observatories,
which is operated by the AURA, Inc. under cooperative agreement with the National Science Foundation} and our
own custom software.  We applied overscan and bias subtraction to all images in the standard way.  Separate
flat field images of the telescope dome white spot were obtained at each observed wavelength, and were
used to correct the detector pixel-to-pixel response variations and to remove the H$\alpha$ filter
transmission response.  The wavelength scale was calibrated using a series of comparison exposures of H$\alpha$
and nearby Neon lines.  The zero-point drift of the wavelength solution was determined from H$\alpha$
calibration exposures interspersed with the object frames during the observations.  We estimate that
the resulting wavelength calibration is accurate to better than 0.1 \AA.  
The RFP instrumental spectral profile measured from the calibration exposures is well fit by
a Voigt function with a Gaussian width of 44 \kms and Lorentzian width of 52 \kms.  The resulting instrumental
FWHM is 117 \kms.  We removed cosmic rays from the images by interpolating over neighboring pixels and used
foreground stars to register all the images to the one at the central velocity, $\Delta$V\,=\,$-$18
\kms.  We scaled the transmission of the RFP scans to a common airmass using the CTIO mean extinction tables,
then applied a Gaussian smoothing kernel to each image to produce a common effective seeing of 2\farcs8\, FWHM.
Finally we computed a coordinate solution (accurate to within 0\farcs2) for each image using USNO$-$A2.0 
Catalogue stars in the field.  

\section{SPECTRAL EXTRACTION AND BACKGROUND SUBTRACTION}

The RFP datacube enables us to extract spectra from any desired point along the blast wave of \s2.
Using snapshots of \s2 at different velocity intervals, we can decide on appropriate locations, 
sizes and orientations for spectral extraction apertures.  This is particularly advantageous in the case 
of \s2, where the mixture of non-radiative and partially radiative shock emission around the remnant (Tuohy \etal 1982,
SKBW91, SRL94) forces us to exercise caution when selecting regions for spectral analysis.

In addition to having H$\alpha$ narrow components contaminated
by radiative recombination, most of the bright lumps and filaments exhibit [N~II] $\lambda\lambda$ 
6548, 6583 emission lines in their spectra.  The broad component widths are large enough to overlap
with these lines, complicating the task of fitting the H$\alpha$ profiles.  Therefore, we focused our
spectroscopic analysis on parts of the SNR which have suffered less deceleration than those studied earlier
by SKBW91 and SRL94.

After trying various aperture selection schemes, we decided that the best option would be to preserve 
a constant number of integrated counts for all of the spectra. This results in H$\alpha$ line profiles with
similar signal-to-noise and facilitates the inter-comparison of their broad-to-narrow ratios.
Since the thickness of the blast wave filaments
is nearly constant throughout the remnant, we chose apertures with constant widths and varying lengths.  
While selecting spectral extraction apertures for \s2, we were careful to avoid the aforementioned
regions showing signs of radiative cooling, namely the bright filaments seen
in the eastern, northern and western portions of the remnant.  We also excluded regions where stars
were present.  The final selection of 14 apertures is
marked in Figure~1 and listed in Table~1.  Each aperture
contains approximately 13,000$\pm$500  counts over the entire line profile, before sky subtraction.
The aperture sizes, locations and position angles were chosen to cover as much of the blast wave
emission as possible while preserving a constant number of counts per aperture and avoiding the
brighter filaments on the east-west sides of \s2.

The RFP spectral extraction consisted of summing the emission within each defined aperture,
for each frame of the datacube.  Although the wavelength varies with position in each FP image,
the variation over each aperture is  $<$ 0.01 \AA, negligible compared to the spectral resolution.
To perform the sky subtraction, we selected a ring of pixels in each frame of the datacube, centered
on the optical axis and passing through the middle of each extraction aperture.  This ensured that
the sky fluxes were extracted at the same wavelength as the emission within each
blast wave aperture.  The ring included all the sky lying off the supernova remnant
which did not fall onto stars or other detectable discrete sources.  We summed
the sky emission along the ring, rejecting pixels deviating more than 4$\sigma$ from the mean value.
We then multiplied the resulting sky spectrum by $N_{pix}(obj)/N_{pix}(sky)$
to obtain a scaled sky spectrum for each of the blast wave apertures. 
Subtracting this spectrum from that of our object yielded the final object spectrum for each
aperture (Figs.~2a$-$2d).  The sky subtraction removes large scale sources of background emission in the
RFP data bandpass, namely night sky H$\alpha$, galactic H$\alpha$ and [N~II] $\lambda$6583,
and the LMC H$\alpha$ and [N~II] $\lambda$6548.   

\section{LINE PROFILE FITS}

The intrinsic SNR Balmer line profiles are well described by the sum of two Gaussian shapes (Ghavamian
1999) with different widths, velocities and strengths.  These are convolved with the instrumental response
function of the Fabry-Perot, a Voigt profile.  Therefore, the resulting line shape is the sum of two Voigt
profiles.  Since the narrow component line is unresolved, it is characterized by two parameters: its central
wavelength and flux.  The broad component, on the other hand, is described by three parameters: its
central wavelength, Gaussian width and flux.  In fitting the sky-subtracted profiles we leave the five
listed parameters free, while fixing the baseline level to zero.

Due to the presence of Lorentzian wings in the extracted line profiles, it is not always
obvious from the data whether the wings of an H$\alpha$ profile are caused by a second
(broad) component or whether they are simply the Lorentzian tail of a single, narrow component.  Since
the broad-to-narrow flux ratio can vary significantly from point to point along the rim of \s2, we 
required a method for quantitatively measuring the likelihood of a broad component detection in each extracted spectrum.
We used the following approach.  First, we fit each H$\alpha$ profile with a single, narrow
Voigt function.  Using parameters output from a $\chi^{2}$-minimized fit, we then added a second
Voigt function and re-computed the best fit.  The initial guesses for the broad component parameters were to
take \bn = 0.5 and to assume an unresolved broad component profile, $\sigma_{G}(b)\,=\,\sigma_{G}(n)$. 

To statistically evaluate the influence of a second Voigt component on the profile fits, we computed
an F-statistic for each profile using the definition of F and the goodness-of-fit parameter
$P(\geq F)$ from Martin (1971) and Band \etal (1996).  In this procedure a large value of $F$ and
small value of $P(\geq F)$ would indicate that 
it is unlikely the superior two-component fit resulted from a random statistical fluctuation.  As 
Table~2 shows, $P(\geq F)$ is on the order of 5\% for all blast wave apertures other than
apertures 1 and 2, indicating that a second, broad component is detected in each H$\alpha$
profile.

\section{BLAST WAVE SPECTRA}

Most of the blast wave around \s2 follows a `faceted' morphology, where relatively straight
sections of the shock front appear joined together at abrupt breaks in orientation angle. This is most
strikingly seen in the northern, southwestern and southeastern edges of the remnant.  The
size of each section reflects the length scale for variations in ambient density around \s2.

The H$\alpha$ profile fits in Figures~2a$-$2d show that the H$\alpha$ broad component is detected in 14 of the 16 spectra.
The fitting results are summarized in Table~2.  The broad component widths and broad-to-narrow ratios vary considerably
with position along the rim.  The faintest blast wave filaments tend to exhibit the
largest broad component widths (with the notable exception of aperture 10), while sections of the blast
wave lying adjacent to filaments of intermediate brightness (such as apertures 4, 13 and 14) tend to exhibit smaller broad
component widths.

Spectra from apertures 1 and 2 at the northern rim are well fit by a single narrow
H$\alpha$ emission line, indicating that the broad component emission either lies beneath the detection
threshold or is entirely absent at these locations.  Interestingly, a broad component is detected in
the aperture 3 spectrum, even though it is part of the same straight, faint filament as
apertures 1 and 2 (Figure~1).  This is likely due to a gradient in preshock neutral density which causes the surface brightness
to increase by nearly 70\% from aperture 1 to aperture 3.  A density gradient is also apparent in the
X-ray surface brightness of the \chan data (RGH03), indicating that the total preshock density also rises
between apertures 1 and 3.  The broad component widths of apertures 3 and 4 and their corresponding
errorbars are consistent with a single broad component width of 840$^{+115}_{-100}$ \kms, nearly 40\%
larger than the widths measured by SKBW91 from the brighter east/west filaments.

Progressing along the western side of \s2, we find that the spectra of apertures 5 and 6 exhibit the largest 
broad component widths ($\sim\,$1000 \kms) seen to date in this SNR.  The faint blast wave bulges out beyond the
brightest western filament (Figure~1).  Since the Balmer emission is produced very close to the blast wave,
the radial separation between two filaments is entirely due to geometric projection.  This
suggests that the brightest western emission marks the location where the forward shock has slowed
due to an encounter with denser material.  This is confirmed by both longslit spectra (SKBW91) and our RFP 
profiles, which indicate that the broad component width of the bright western filament is only half that of 
apertures 5 and 6.  The broad component profiles of apertures 5 and 6 are consistent with a single 
FWHM of 985$^{+210}_{-160}$ \kms and \bn=\,0.54$\pm$0.09.  These are the largest broad component
widths reported so far in \s2.

The southern edge of \s2 reveals some of the best examples of the `faceted' rim morphology mentioned
above.  A noticeable brightening occurs in the filament covered by apertures 7$-$9.  The {\it Chandra}
image of \s2 shows that the enhanced H$\alpha$ is accompanied by a sharp rise in X-ray emission 
(RGH03, HGRS03).  This indicates a higher total preshock density at the regions covered by
apertures 7$-$9.  All three apertures
show similar broad component widths and broad-to-narrow ratios (Table~2).  Combining the spectra from the three apertures, 
we obtain a broad FWHM of 805$^{+140}_{-115}$ \kms\,and \bn=\,0.49$^{+0.07}_{-0.06}$, similar to values seen in 
apertures 3 and 4.  The broad-to-narrow ratio and broad component width vary considerably from aperture 9 to 12,
suggesting substantial variations in shock velocity, preshock density and possibly preshock neutral fraction.  The
component widths of the aperture 13 and 14 spectra are nearly equal to the values reported by SKBW91 for the
bright, partially radiative eastern filament.  

Unlike the other shocks in our study, the aperture 15 and 16 filaments are seen well inside 
the blast wave rim (Figure~1).  The filaments
sampled by these apertures are of similar surface brightness, yet the broad component width and broad-to-narrow
ratio of the aperture 15 spectrum are more than twice as large as the aperture 16 values.  The
broad component shift in aperture 16 is 160 \kms relative to the narrow component,
suggesting that the H$\alpha$ emission in this aperture is dominated by partially face-on shocks.
The {\it Chandra} image of \s2 shows that there are in fact three nested blast wave
shocks in the southwest region (RGH03).  By comparing the optical images with the {\it Chandra} image we
have determined that two of these nested shocks are sampled by aperture 15.  In that case,
variations in the line-of-sight bulk velocity from inter-filament emission could contribute substantially
to the broad component width of the aperture 15 spectrum.

\section{SHOCK MODELS}

To model the Balmer line profiles, we used the one-dimensional, plane-parallel shock
code described in Ghavamian (1999) and GRSH01 to compute a grid of numerical models.   The
code calculates the density and temperature of electrons, protons and hot neutrals behind a Balmer-dominated shock.
Broad-to-narrow flux ratios are computed by a Monte Carlo simulation which follows the excitation of H$\alpha$ photons
in the broad and narrow components.  The radiative conversion of Ly $\beta$ photons into H$\alpha$ (Ly $\beta$ trapping)
is also followed in the simulation.  At the shock speeds relevant
to \s2 ($\lesssim\,$1500 \kms) the proton$-$H ionization and excitation rates are $<$ 1\% of the
electron rates. Therefore, all direct collisional ionization/excitation processes are dominated by the electrons.
The shock velocity, fractional electron-ion equilibration \tetp at the shock front by plasma turbulence,
and preshock neutral fraction of H are taken as free parameters.  The quantity \tetp
varies from $m_{e}/m_{p}$\,($\sim$\,0)
for zero equilibration at the shock front to \tetp=\,1 for full, prompt equilibration
at the shock front. (In this work we have chosen to describe our results in terms of \tetp rather than
the related quantity $f_{eq}$ used in earlier works (GRSH01, Ghavamian \etal\, 2002), to maintain
a consistent notation between this paper and the companion work by RGH03).

In addition to the shock code improvements listed in Ghavamian \etal\, (2002), we have included
the energy lost by electrons to collisional excitation and ionization.  Loss terms are included for collisional ionization 
of H and He, and for Ly$\alpha$ and two-photon continuum excitation of H and He.   
Clearly, ionization losses are most important at the lowest shock speeds,
lowest equilibrations and highest preshock neutral fractions, where the electron temperature close to the shock
front is lowest.  In these cases $T_{e}$ can initially decline with position
behind the shock as excitation and ionization losses exceed energy gain by Coulomb collisions. 
Subsequently as the ionization fraction rises, $T_{e}$ also rises over a length scale comparable
to the thickness of the H ionization layer as Coulomb collisions come to dominate
the energy balance.  In models with shock speeds $\sim\,$500 \kms, \tetp\,$\lesssim$\,0.4 and
preshock neutral fractions $\gtrsim$ 0.9, we estimate that ionization losses produce a 5\%$-$7\% increase in the 
broad-to-narrow ratios relative to models that neglect these losses.

Using the method outlined in GRSH01, we computed a grid of shock models for
spectra extracted from the blast wave rim.   We matched each grid point in \tetp with the
shock velocity \vs required to match the observed FWHM of the broad component line.  Using this
set of parameters (\vs, \tetp), we calculated \bn for a range of assumed
preshock neutral fractions \fneut, then compared the results with the observed \bn.
However, before producing the grid we decided to combine spectra of regions
exhibiting consistent broad-to-narrow ratios and broad component widths (to within the errors). 
The main purpose of combining the spectra was to define large enough apertures along the blast wave to
obtain adequate statistics in our parallel analysis of {\it Chandra} data (RGH03).  The five combined regions
are labeled 3+4, 5+6, 7+8+9, 11+12 and 13+14 in reference to the original 14 apertures (c.f. Table~2).
The combined apertures correspond to X-ray spectral extraction regions X1$-$X5 from RGH03.
After combining the spectra, we re-fit the line profiles in the manner described earlier to estimate
the broad component FWHM and broad-to-narrow ratios (Table~3).  The limits on \vs implied by the 
combined aperture broad component widths are listed in Table~4 for the cases \tetp\,=\,\memp and \tetp\,=\,1.

\section{COMPARISON BETWEEN OBSERVATIONS AND SHOCK MODEL PREDICTIONS}

From Figure~3 it is obvious that there is a major discrepancy between the observed and predicted broad-to-narrow ratios:
the observed ratios lie as much as 50\% below the smallest modeled values.   Clearly, some aspect of the shock physics
is not included in the models.   It seems likely that our models overpredict the broad-to-narrow ratios because they
underpredict the narrow component H$\alpha$ emission.  Narrow component H$\alpha$ emission can be
enhanced by processes not included in our Balmer-dominated shock models.  Although our calculations include the enhancement
of narrow component H$\alpha$ from Ly$\beta$ trapping within the shock, other possible sources of narrow Ly$\beta$, such
as other shocks in the SNR, are ignored. In addition, if the preshock gas is heated in a precursor it may produce enough 
collisionally excited H$\alpha$ to lower the observed broad-to-narrow ratios.

One way of bringing the model broad-to-narrow ratios into agreement with the observations would be
to forgo the Monte Carlo simulation altogether and simply compute the integrated H$\alpha$ flux assuming
a negligible Ly$\beta$ optical depth in the broad component (case A conditions) and a high Ly$\beta$ optical depth 
in the narrow component (case B conditions) (Chevalier, Kirshner \& Raymond 1980, SKBW91, Ghavamian 1999).
However, we find that although the Ly$\beta$ optical depth can
be large enough at line center to make case B a valid approximation for the narrow component, it is
generally not small enough at broad component line center to allow a case A approximation for the broad component.
For example, $\tau_{Ly \beta}(narrow)$ $\sim$ 7 and
$\tau_{Ly \beta}(broad)$ $\approx$ 0.2 for \vs \,$\sim$ 1000 \kms and \fneut = 0.9. Smaller neutral fractions
can lower $\tau_{Ly \beta}(broad)$, but not enough to ignore Ly$\beta$ trapping in the broad component.
Therefore, usage of the case A/case B approximations in the computation of broad-to-narrow ratios requires physical conditions
not met in practice.

The limiting assumptions from above can be made more plausible if we include absorption of narrow Ly$\beta$ photons produced
in other parts of the remnant.  For a given shock, this would effectively increase $\tau_{Ly \beta}(narrow)$.   
Absorption of nonlocally produced Ly$\beta$ is more important for a remnant like \s2 which is completely surrounded
by Balmer filaments than remnants with only partial shells of Balmer emission.  
A Ly$\beta$ photon may propagate all the way to the far side of the remnant without
being absorbed.  Once the photon enters the shock structure on the far side of the remnant, it will most likely
be converted into narrow H$\alpha$ as it propagates upstream.  From the Monte Carlo shock models we estimate
that for shock speeds between 600 and 1000 \kms, preshock neutral fractions greater than 0.1 and all values
of \tetp, less than 8\% of all narrow component Ly$\beta$ photons excited behind the \s2 blast wave escape downstream.
If all of these photons are absorbed in other parts of the blast wave and are converted into narrow H$\alpha$,
the broad-to-narrow ratios be lowered by at most 20\% from their intrinsic values.  This is less than half
the amount required to match most of the observed broad-to-narrow ratios with the smallest model values. 
Although the Ly$\beta$ luminosity of the 
partially radiative shocks in \s2 is nearly twice as large as that of the fainter, pure non-radiative shocks, their
narrow Ly$\beta$ flux is offset by the significantly smaller surface area of the remnant covered by the partially radiative shocks.
Therefore, we now consider the possibility that most of the excess narrow component H$\alpha$ is produced ahead of
the blast wave, in a precursor.

\section{EXPLAINING THE BROAD-TO-NARROW RATIOS}

When ionizing photons or energetic particles produced behind a non-radiative shock (such as cosmic rays or fast neutrals) 
cross upstream, they generate a layer of heated preshock gas known as a precursor.  The
characteristic size $d_{p}$ of the precursor depends on the mean free path traveled by the photon or particle
before depositing its energy into the upstream gas.  This energy is shared between electrons and ions via Coulomb collisions, which for
the preshock temperatures inferred from spectroscopic observations (12,000~K$-$40,000~K, Hester, Raymond \& Blair 1994, SRL94, GRHB00) 
leads to $T_{e}\,=\,T_{i}$ ahead of the shock.  Collisional excitation by electrons within the precursor can produce
observable line emission (Fesen \& Itoh 1985, GRHB00).  The detectability of such emission depends
on the thickness, temperature and ionization structure of the precursor.  Therefore, we must constrain these parameters 
in order to investigate the influence of precursor emission on the observed broad-to-narrow ratios.  

A constraint on the precursor thickness can be obtained from the optical images of \s2.  If the precursor 
is spatially resolved, it should appear as a layer of diffuse H$\alpha$ emission extending ahead of the Balmer-dominated
filaments (GRHB00).  Since the shock front continually overruns the heated ambient gas, the diffuse precursor emission
should drop sharply behind the Balmer filaments.  A narrow band image of \s2 acquired before the RFP observations does show
marginal evidence for diffuse emission above the local background, extending a few arcseconds ahead of the northern limb.
However, elsewhere in the remnant the surrounding sky does not show any obvious diffuse emission above the background 
LMC level.  There is also no clear evidence of diffuse emission ahead of most of the filaments in the RFP scans,
aside from the emission mentioned earlier ahead of the northern rim.

The remaining possibility is that the precursor emission is produced on small scales, comparable to or less than the
Balmer filament thickness.  
One method of constraining the thickness of such a layer is to examine the RFP images for radial stratification between shocks
observed at pure broad component velocities and shocks observed at narrow component velocities (e.g., less than the instrumental
width of 117 \kms).  If pure broad component filaments appear inside of pure narrow component filaments, this
would be nominal evidence for emission ahead of the blast wave on scales slightly larger than the Balmer filament thickness.
Performing this test with the RFP images, we find no obvious spatial shift between the narrow and broad emission.  Therefore,
the layer of excess narrow component emission is restricted to a thickness $\lesssim\,$3$^{\arcsec}$ (the seeing FWHM).  This
restriction is important because it allows us to constrain the type of precursor producing the excess H$\alpha$ emission.

Before examining the various precursor mechanisms in detail, we must consider the relationship between the spatial
resolution of the detector and the relative contributions of the shock and precursor emission from the Balmer-dominated
filaments.  At a given pixel scale, the spatial resolution of a detector imaging an LMC remnant is over an order of magnitude
worse than that of a detector imaging a Galactic remnant.  This can make the spatial separation of the precursor and shock
emission much more difficult for an LMC remnant like \s2.  If the thickness of the precursor $d_{p}$ and the thickness of the postshock
H ionization zone $d_{s}$ are both smaller than the projected size of the seeing FWHM, it will be nearly impossible to subtract the precursor
emission from that of the shock in a Balmer-dominated filament.  Furthermore, the ratio of precursor to shock surface brightness $I_{p}/I_{s}$
depends on the viewing angle to the shock. If $d_{p}$ $>$ $d_{s}$, but both are smaller than the projected size of the seeing FWHM, 
$I_{p}/I_{s}$ can become significant for edge-on filaments, even if the ratio is small at face-on positions.

Assuming a spherical geometry and that the radius R of the remnant extends from the remnant center to the blast wave
shock front, the edge-on emission is related to the face-on emission through a factor $sec(\theta)\,=\,L/d$, where $L$ is the
depth of emission viewed edge-on and $d$ is the larger of the layer thickness and the projected size of the atmospheric seeing.
If $d_{s}$ and $d_{p}$ are both smaller than the projected seeing, the ratio is
\begin{equation}
\left(\frac{I_{p}}{I_{s}}\right)_{e}\,=\,\left(\frac{I_{p}}{I_{s}}\right)_{f}\,
\frac{\sqrt{(R\,+\,d_{p})^{2}\,-\,R^{2}}}{\sqrt{R^{2}\,-\,(R\,-\,d_{s})^{2}}  }\,\,\approx\,\,
 \left(\frac{I_{p}}{I_{s}}\right)_{f}\,\left(\frac{d_{p}}{d_{s}}\right)^{1/2}
\label{ipis}
\end{equation}
where we have used $R\,\gg\,d_{s}$, $d_{p}$ to simplify the expression. The ratio of precursor to shock emission for the face-on
position is 
\begin{equation}
\left(\frac{I_{p}}{I_{s}}\right)_{f}\,=\,\frac{n_{e}\,q_{H\alpha}(T_{e})\,d_{p}}{0.2\,V_{S}  }
\label{ipisf}
\end{equation}
where q$_{H\alpha}$ is the H$\alpha$ collisional excitation rate in the precursor, $T_{e}$ is the electron temperature
in the precursor,  0.2 is the number of H$\alpha$
photons excited per H atom, $V_{S}$ is the shock speed and $n_{e}\,=\,f_{H^{+}} n_{0}$ is the preshock electron density.
Using Equations \ref{ipis} and \ref{ipisf} and our limits on the shock parameters, we can first calculate $(I_{p}/I_{s})_{f}$,
then use Equation \ref{ipis} to find $(I_{p}/I_{s})_{e}$, the ratio expected for the filaments.   Although some spatial variation of
$T_{e}$ and $f_{H^{+}}$ is expected within the precursor (Boulares \& Cox 1988, GRHB00), we take these quantities to be
constant for the crude estimates in this section.

The amount of precursor emission required to match the observed broad-to-narrow ratios depends on the assumed
shock parameters.  For an observed broad-to-narrow ratio
(\bn)$_{o}$ and specified model broad-to-narrow ratio (\bn)$_{m}$, the implied ratio of precursor 
to shock emission is
\begin{equation}
\frac{I_{p}}{I_{s}}\,=\,\frac{   ((I_{B}/I_{N})_{m}/(I_{B}/I_{N})_{o})\,-\,1    }{(I_{B}/I_{N})_{m} + 1}
\label{precs}
\end{equation}
where we have assumed that the observed narrow component flux is given by the sum of fluxes from the shock
and the precursor.  From Figure~3, the smallest broad-to-narrow ratios predicted by our models lie in the
range 0.9$-$1.1.  The observed broad-to-narrow ratios lie in the range 0.4$-$0.75.  Therefore, using Equation \ref{precs},
the edge-on ratio of precursor to shock emission must lie in the range 0.1 $\leq$ $(I_{p}/I_{s})_{e}$ $\leq$ 0.9 to 
resolve the broad-to-narrow ratio discrepancy.  Therefore, matching the smallest (\bn)$_{o}$ with (\bn)$_{m}$ 
can require nearly equal contributions to the H$\alpha$ flux from the precursor and shock.

For the purpose of obtaining order-of-magnitude estimates of $I_{p}/I_{s}$, the main distinction
between the precursor scenarios lies in the estimated values of $d_{p}$ and $T_{e}$.  We consider three scenarios below:

(1) {\it Photoionization Precursor$-$} The gas behind all non-radiative shocks produces ionizing radiation
which propagates far upstream and heats the preshock gas (Draine \& McKee 1993, SRL94, GRHB00).
The dominant source of ionizing photons is He~II $\lambda$304 (Ly $\alpha$) line emission, which can heat
the preshock gas enough to produce observable optical emission ($T_{e}\,\sim\,$12,000~K$-$20,000~K, GRHB00) but 
not enough to establish ionization equilibrium.
The characteristic size of the photoionization precursor is one mean free path of a He~II $\lambda$304 photon,
$d_{304}\,=\,(n_{H^{0}}\,\sigma_{304}$)$^{-1}$ (SRL94, GRHB00).  
Assuming a total preshock density $n_{0}\,\lesssim$ 0.5 cm$^{-3}$ (as constrained by {\it Chandra} spectral fits of RGH03) and
neutral fractions lower than 0.9, we obtain an angular size of $>$\,7\arcsec\, for the precursor, sufficiently large to be
resolved in the RFP images.  However, as mentioned earlier, there is no clear evidence of such emission ahead of the
Balmer filaments and above the LMC background.  Therefore, we conclude that although a He~II $\lambda$304 is undoubtedly present, it 
does not contribute enough to the H$\alpha$ filament emission to account for the anomalously low broad-to-narrow ratios.

(2) {\it Fast Neutral Precursor$-$}  Fast neutrals on the tail of the distribution can cross upstream and deposit energy into the
preshock medium via elastic collisions and charge exchange with the preshock ions (Hester, Raymond \& Blair 1994, SRL94, Lim \& Raga 1996).  
Although there is considerable uncertainty regarding the energy 
transfer, existing studies suggest that heating in the precursor can be substantial (SRL94, 
Lim \& Raga 1996), with temperatures in excess of 20,000~K easily attained.  However, the
thickness of the fast neutral precursor is on the order of the charge exchange mean free path, 
$d_{p}\,=\,d_{cx}\,\sim\,$10$^{15}/n_{0}\,\approx\,$2$\times$10$^{15}$ cm for the deduced preshock parameters.  We estimate
the electron temperature in the precursor from measured narrow component widths of SRL94.  Assuming purely thermal line broadening 
and \tetp\,=\,1 in the preshock gas, the 30 \kms$-$45 \kms 
narrow component widths quoted by SRL94 imply an electron temperature of 20,000~K$-$40,000~K for the precursor.  The
collisional excitation rate coefficient $q_{H\alpha}$ rises steeply with $T_{e}$ over this temperature range.  From
the Case B calculation of Aggarwal (1983), we estimate $q_{H\alpha}\,\approx\,$(0.1$-$3.5)$\times$10$^{-10}$ cm$^{-3}$
s$^{-1}$.  We adopt $n_{0}\,\approx\,$0.5 cm$^{-3}$ and take
the preshock ionization fraction $f_{H^{+}}$ to lie between 0.2 and 0.9.  Photoionization by He~II $\lambda$304 and
X-ray photons from the reverse shock would make ionization fractions smaller than 0.2 unlikely, while fractions larger
than 0.9 would make the Balmer filaments too faint to observe.  We assume $V_{S}\,=\,$900 \kms using the average speed
listed in Table~4 for the least decelerated portions of the blast wave.

Inserting the estimates from above into Equation \ref{ipisf}, and assuming that $d_{p}\,\approx\,d_{cx}$, we
obtain $(I_{p}/I_{s})_{f}\,\lesssim$\,0.1.  Therefore, explaining the full range of broad-to-narrow ratios
measured in the filaments remnant would require $(d_{p}/d_{s})^{1/2}\,\gtrsim\,$10.  However, for the shock
speeds found in \s2, $d_{s}\,\sim\,d_{cx}\,\sim\,d_{p}$, so there is little enhancement in the ratio of precursor to
shock emission at edge-on viewing angles, $(I_{p}/I_{s})_{e}\,\approx\,(I_{p}/I_{s})_{f}$.
Therefore, narrow H$\alpha$ emission from a fast neutral precursor might contribute
to the blast wave spectra of \s2, but not enough to resolve the discrepancy in all the broad-to-narrow ratios.

(3) {\it Cosmic Ray Precursor$-$} Cosmic rays are accelerated by the shock when they scatter from self-generated
Alfven turbulence upstream.   Since the size of the cosmic ray precursor depends on the poorly known diffusion coefficient of
cosmic ray ions, we can only place an upper limit on its size by requiring that the precursor layer be thin enough to avoid
ionization equilibrium (otherwise no neutral H would remain to produce the Balmer-dominated emission).  This
requires $d_{p}\,<\,v_{S}/n_{e}\,q_{i}$, where $q_{i}$ is the ionization rate coefficient in the precursor (Hester, Raymond
\& Blair 1994, SRL94).
Using $q_{i}(40,000~K)\,\approx\,$3$\times$10$^{-10}$ cm$^{3}$ s$^{-1}$ (Janev \etal\, 1987), we obtain $d_{p}\,<\,$6$\times$10$^{17}/f_{H^{+}}$.
For $f_{H^{+}}\,\gtrsim\,$0.2, $d_{p}$ is smaller than the projected size of the seeing FWHM.  Inserting this upper limit $d_{p}$ into Equation
\ref{ipisf} and using $d_{s}\,\approx\,$10$^{15}/n_{H^{+}}$ (the thickness of the H ionization zone behind the shock), we find that
$(I_{p}/I_{s})_{f}$ can reach values as large as 5.8.  According to Equation \ref{ipis}, this ratio is enhanced by another factor
$(d_{p}/d_{s})^{1/2}\,\sim\,$25 at edge-on viewing angles.   Therefore, even if $d_{p}$ were several orders of magnitude
smaller than the value set by the ionization length argument, the narrow H$\alpha$ emission from a cosmic ray precursor would
still be able to lower the intrinsic broad-to-narrow ratios down to the observed levels.

If a cosmic ray precursor exists in \s2, it should generate optical forbidden line emission.
Using the precursor density, temperature and size from above, we can place an upper limit on the [S~II] and [N~II]
surface brightness.  If the preshock ionization fraction is 0.2 $\lesssim\,f_{H^{+}}\,\lesssim$ 0.9, and
$f_{S^{+}}\,=\,$1 and log(S/H) + 12 = 6.87 (Russell \& Dopita 1990), then using the collision strengths of 
Cai \& Pradhan (1993) and assuming the low density limit, we find that the ratio of precursor [S~II] (6716 + 6731) 
to precursor H$\alpha$ is $\lesssim$ 5$\times$10$^{-3}$, too faint for detection.  Similarly, if we take
$f_{S^{+}}\,\approx\,f_{H^{+}}$ and use log(N/H) + 12 = 6.55 (Russell \& Dopita 1990), the ratio of
[N~II] ($\lambda$6548+6583) to H$\alpha$ is $\lesssim$ 3$\times$10$^{-4}$ in the precursor.   At a
temperature of 40,000~K, the H collisional excitation rate is more than an order of magnitude larger than the
rate at lower temperatures (12,000~K$-$15,000~K, as found in a photoionization precursor, GRHB00).
On the other hand, the forbidden line excitation rates do not increase as rapidly between 12,000~K and 40,000~K.
Therefore, the Balmer line emission dominates over forbidden line emission in the precursor.  This is consistent
with the lack of detection of [S~II] emission from the pure Balmer-dominated filaments in our narrow band images
of \s2.

Given the uncertain contribution of precursor emission to the narrow component in \s2, we refrain from
further attempts to use the observed broad-to-narrow ratios to estimate \tetp\, and \vs.  Instead, we
have combined the proton temperatures and range of shock speeds implied by broad component widths around the rim 
with electron temperature measurements from {\it Chandra} data to estimate \tetp\, and \vs (RGH03).  In the
rest of this paper, we use these estimates to draw conclusions regarding the age and explosion energy of \s2. 

\section{Evolutionary Models of \s2}

The large radius ($\sim\,$10 pc) and moderate blast wave speed ($\sim$400$-$1000 \kms) of \s2 suggest that this remnant 
is well into the Sedov stage of evolution.  One way to test this conclusion is to compare scale quantities predicted
by Sedov models, such as the characteristic radius, with the observed radius of the supernova remnant.
In the case of a SNR in a uniform density medium and an exponential ejecta profile (as expected for Type Ia explosions;
see Dwarkadas \& Chevalier 1998, Wang \& Chevalier 2001), the blast wave initially evolves through an ejecta-dominated phase.
It then enters the Sedov stage when its radius has grown to a characteristic size $R_{ST}$, given by
\begin{equation}
R_{ST}\,=\,2.2\,\left(\frac{M_{ej}}{M_{ch}}\right)^{1/3}\,n_{0}^{-1/3}\, \, \rm pc
\label{rst}
\end{equation}
(Wang \& Chevalier 2001), where $M_{ej}$ is the ejected mass, $M_{ch}$ is the Chandrasekhar
mass (1.4 $M_{\odot}$) and $n_{0}$ is the density of the ambient medium.  
Taking a range 0.3\,$\lesssim\,n_{0}\,\lesssim$1.4 from fits to the combined aperture {\it Chandra} blast 
wave spectra (RGH03; also see Table~4) and assuming $M_{ej}\,=\,M_{ch}$, we find $R_{ST}\,\lesssim$3.3 pc.  This is less than half the
average blast wave radius of \s2; therefore the remnant should be well into the Sedov-Taylor phase of evolution.

Two fundamental parameters of interest in \s2 are the age $\tau$ and explosion kinetic energy $E_{51}$.  
We assume that azimuthal variation in \vs and $R$ result from azimuthal variations in ambient density.  A reasonable
first approximation for modeling this dependence is to describe the blast wave evolution in each spectral extraction
aperture in terms of its own Sedov solution.  In that case, $E_{51}$ and $\tau$ are identical everywhere along the
blast wave rim, with $n_{0}$ varying from section to section.
In that case, the equations governing the evolution of the remnant predict $V_{S}\,=\,\alpha\,$R,
where $\alpha\,\equiv\,\frac{2}{5\tau}$.  Furthermore, inverting the Sedov-Taylor relation (Sedov 1959) for R(t) yields
\begin{equation}
n_{0}(R) = 3.1\times10^{-3}\,\tau^{2}\,E_{51}\,R^{-5}\, \, \hskip 0.11 in \rm cm^{-3},
\label{nvsr}
\end{equation}
where n$_{0}$ is the preshock H density and $\tau$ is measured in years.  Here, the total preshock
density in the Sedov equation is related to the preshock H density (measured from the X-ray spectra) by a factor of 1.4.
Fitting $V_{S}(R)$ and $n_{0}$(R) to the data yields $\tau$ and $E_{51}$.  To perform these fits,
we utilized the values of \tetp determined by RGH03 to calculate the appropriate shock speeds consistent with the
broad component widths found in the combined aperture regions (see Table~4).  We also used $n_{0}(R)$ determined from the
X-ray models of RGH03.  Since GRSH01 and RGH03 found evidence
that \tetp and \vs are inversely related, the faster shocks in \s2 will have shock speeds closer to the \tetp\,=\,0 predictions,
while the slower shocks will have speeds closer to the \tetp\,=\,1 predictions.  The assumption that the combined aperture shocks
are either all unequilibrated or fully equilibrated is not strictly valid; however, the results of our fits to $V_{S}$(R) and $n_{0}(R)$ 
do not change significantly if \tetp is set to either extreme for all the shocks.

The analysis outlined above requires an estimate of R for each aperture, which in turn requires an estimate of the expansion center.
Therefore, the following procedure was implemented: first we used a range of expansion center coordinates to calculate a grid
of fits for $V_{S}$(R).  After choosing the expansion center coordinates which yielded the smallest $\chi^{2}$
in the grid, we then computed the 1$\sigma$ error box about that best fit center (see Figure~1).  Next, we independently performed the same
procedure for n$_{0}$(R), obtaining an independent estimate of the expansion center and a 1$\sigma$ error box on that fit 
(also marked on Figure~1).   We chose the central pixel from the overlap region between the $V_{S}$(R) and $n_{0}$(R) error boxes
and designated the coordinates of that pixel ($x_{cen}$, $y_{cen}$) to be the expansion center of the remnant.  Finally,
returning to the grids in $V_{S}(R)$ and $n_{0}$(R), we use the fits at the expansion center ($x_{cen}$, $y_{cen}$) to
estimate $\tau$ and $E_{51}$.

In Figure~4 we show the fitted blast wave speeds vs. radii, derived using our best estimate for the expansion center
(located at $\alpha(2000)\,=\,$05$^{\rm h}$05$^{\rm m}$42\fs 3, $\delta(2000)\,=\,-$67$^{\circ}$52\arcmin 41\farcs3).
Here we have ignored the uncertainty in blast wave radius in our images.  Since the distance to the LMC is uncertain by
as much as 13\% (Gibson 2000), we have included a factor $D_{50}$ (the distance in units of 50 kpc) when quoting the 
fitted parameters. The slope $\alpha$ of the
fit is (89.6$\pm$5.9)\,$D^{\,-1}_{\,50}$ \kms pc$^{-1}$, which yields $\tau$\,=\,(4360$\pm$290)\,$D_{\,50}$ years for \s2.  This age is less than
half the value quoted by SKBW91, whose analysis was based on broad component width measurements of the slower, brighter
Balmer-dominated shocks.  We also note that an age of (4360$\pm$290)\,$D_{\,50}$ years is in good agreement with the age of 4700
years calculated from fits to {\it ASCA} data by HHK98.  

Using an expansion center fixed at ($x_{cen}$, $y_{cen}$) and our estimated remnant age, the fit
to n$_{0}$(R) (shown in Figure~5) yields an explosion kinetic energy of $E_{51}$\,=\,(0.37$\pm$0.06)\,$D^{\,5/2}_{\,50}$.
The two main sources of uncertainty lie in the distance to the LMC and the FWHM measurement of the broad component 
width (which affects $\tau$ through \vs).  Allowing for the distance uncertainty to the LMC and the statistical error,
$E_{51}$ lies between 0.2 and 0.6, less than half the value
computed by HHK98.  The main reasons for the difference are that HHK98 approximated \s2 as a sphere
10.4 pc in radius (despite the sizable variation of R with azimuth) and fit the global X-ray spectrum
of both the blast wave and metal rich ejecta to generate Sedov models for the remnant.  Our optical study
utilizes velocities, densities and radii from multiple sections of the blast wave; therefore, it provides a much
more direct and accurate estimate of the explosion energy than the models of HHK98.
However, our derived explosion energy for \s2 is low compared to values predicted by 1-dimensional
(H\"{o}flich \& Khokhlov 1996) and 3-dimensional (Reinecke \etal\, 2002) hydrodynamical models of Type Ia
SNe.  The reason for the difference is not clear; one possible explanation is that some of the kinetic 
energy imparted to the ejecta in \s2 was lost during an earlier epoch when, unlike now, cosmic ray acceleration
was efficient.

The expansion center overlap region in Figure~1 is shifted
slightly to the east of the apparent center. This is consistent with a supernova explosion
in a medium with an azimuthal density gradient.  A higher than average density has
decelerated the blast wave on the entire east side of the remnant, while recent encounters with H~I
clouds on the east and west sides have caused a strong brightening in the optical emission.  The
presence of forbidden line emission in these filaments indicates that these shocks are
becoming radiative.  In contrast, the blast wave is expanding most rapidly
along the north and southwest rims of the remnant, where the density is lower than average.
The fact that the preshock densities follow an R$^{-5}$ relation so well implies that despite
the large differences in shock speed between the east rim of \s2 and the rest of the remnant, the
kinematics of the blast wave are consistent with Sedov evolution.   

One interesting property of \s2 is that although the partially radiative filaments on the east rim are 
nearly twice as bright as the filaments covered by apertures 13 and 14, they all lie at the nearly the same radius
and have similar broad component widths (as verified by our own data and that of SKBW91). 
The preshock gas in these partially radiative, [N~II]- and [S~II]-emitting clouds is nearly twice as dense as 
that of the shocks in apertures 11+12 and 13+14.  
If the densities of these clouds were plotted on Figure~5, they would lie 
well above the fitted curve for $n_{0}(R)$.  This would reflect a significant departure from the Sedov description
in the brightest filaments, and would suggest that while the shocks in apertures 11+12 and 13+14 
are moving through a moderate, large scale density gradient, the brightest filaments
are shocks driven {\it recently} by the blast wave into stronger, more discontinuous jumps in ambient density.

In the above analysis we have assumed that the blast wave contained within each combined aperture has propagated through
a constant density ISM during the course of the remnant's evolution.  This is a reasonable assumption
since the accumulated weight of evidence suggests that \s2 was produced by an exploding white dwarf (Type Ia SN).
However, we can also determine the age and explosion energy of \s2 under the assumption that
the remnant was produced by a Type Ib or Type II SN inside of a low density cavity.  In that case,
the bubble would have been evacuated by a strong wind from a massive ($\gtrsim$8 M$_{\odot}$) prognitor star.  The 
density profile of the cavity would follow a power law $\rho(R)$\,$\propto$\,$F^{- \omega}$, where $\omega$\,$<$\,0 (Cox \& Franco
1981, Franco \etal\, 1991).  Taking $\omega$\,=\,$-$2 (Franco \etal\, 1991), the relationship between
\vs and R becomes \vs\,=\,$\case{2}{7}\,\case{R}{\tau}$. Using Equation (14) of Cox
\& Franco (1981), the coefficient of R$^{-5}$ in Equation \ref{nvsr} is multiplied by 
(1 $-$ 0.239\,$\omega$)/(3 $-$ $\omega$).  The R$^{-5}$ proportionality in Equation \ref{nvsr} remains unchanged because
the R$^{2}$ dependence of $n_{0}(R)$ is already contained in our measurement of the current density from the X-ray
data.  

Taking the fitted slopes for $V_{S}(R)$ and $n_{0}(R)$ quoted in the previous subsection, we estimate
that for the case $\omega$\,=\,$-$2 the age of \s2 is $\tau$\,=\,(3100$\pm$200)$D_{50}$ yr and 
$E_{51}$\,=\,(0.33$\pm$0.05)$D_{50}^{\,5/2}$.  Thus both the implied explosion energy
and the estimated age of the remnant are slightly lower when the blast wave propagates through a cavity.  This result
is understandable because less energy and less time is required to expand the SNR to its current radius if
the blast wave has only recently encountered the walls of the cavity.  Although steepening the
density profile to $\omega$\,$<$\,$-$2 would lower the blast wave energy still further,
an asymptotic limit is quickly reached where $E_{51}\,\rightarrow\,$0.27\,$D_{50}^{\,5/2}$ as $\omega\,\rightarrow\,-\,\infty$.

The assumption that each section of the \s2 blast wave follows its own Sedov evolution may be questioned
on the grounds that pressure gradients transverse to the blast wave could cause one section to affect the evolution
of another section.  In the general case, the blast wave should be modeled using a multidimensional hydrodynamic
description.  However, as indicated in Table 4, only the shock parameters and radii from region 13+14 differ significantly from 
those of the remaining regions.  Therefore, the strongest pressure differences should arise between regions on
opposite sides of \s2.   The relevant question, then, is to determine how the time $\tau_{cross}$ required for a sound
wave to travel across the remnant interior compares to the age $\tau$ of the remnant.  Using the self-similar relations of 
Cox \& Franco (1981) to compute the sound travel time as a function of average interior temperature, we then integrate
the relation over radius to obtain $\tau_{cross}$.  Taking the current blast wave speed to be 1000 \kms\, and the
diameter of the remnant to be 20 pc, we find that for a remnant age $\sim\,$4000 years (consistent with the diameter
of \s2 and that the ejecta have all passed through the reverse shock as found by HGRS03), $\tau_{cross}/\tau\,\gtrsim\,$3.
This suggests that during the current epoch at least, it is possible for parts of the blast wave on opposite sides 
of the remnant to evolve independently in the manner we have assumed.  

\section{LIMITS ON COSMIC RAY ACCELERATION PARAMETERS}

We can use our derived limits on the precursor thickness, preshock density, shock speed and explosion energy
to constrain cosmic ray acceleration parameters for \s2.  These parameters are the
cosmic ray diffusion coefficient, $\kappa$, the maximum cosmic ray energy $E_{TeV}$ and the roll-off frequency
$\nu_{roll}$ of the synchrotron emission spectrum.   In the limit where $\kappa$ scales linearly with particle energy
(the Bohm limit), the maximum cosmic ray energy is $E_{TeV}\,\approx\,$1.1\,($B_{0}/3 \mu$G) ($E_{51}$\vs/$n_{0}$)$^{1/3}$
(Draine \& McKee 1993), where $B_{0}$ is the preshock magnetic field strength and \vs is in \kms.  Using our shock parameter 
estimates from the previous section, we find $E_{TeV}\,\leq\,$11.3 ($B_{0}$/3 $\mu$G).  The roll-off frequency, defined as the
frequency where the synchrotron power law emission from cosmic rays begins to exponentially decline, is given
by $\nu_{roll}\,=\,$1.4$\times$10$^{13}$\,$E_{TeV}^{\,2}$\,($B_{0}$/3 $\mu$G) Hz (Reynolds 1998).  Inserting our
estimate for $E_{TeV}$, we obtain $\nu_{roll}\,\leq\,$1.8$\times$10$^{15}$\,($B_{0}$/3 $\mu$G) Hz, corresponding
to a wavelength $\lambda_{roll}\,\geq\,$1700\,(3 $\mu$G/$B_{0}$) \AA.  Furthermore, the cosmic ray diffusion coefficient
is related to the precursor thickness via $d_{p}$\,=\,$\kappa$/\vs (Draine \& McKee 1993).  This gives an upper
limit $\kappa\,<\,$5.4$\times$10$^{25}/f_{H^{+}}$ cm$^{2}$ s$^{-1}$ for cosmic rays in \s2.

In an independent study of \s2, Hendrick \& Reynolds (2001) extrapolated the observed radio synchrotron flux of \s2 to
the X-ray regime in order to estimate the maximum cosmic ray energy allowed by the data. They found $E_{TeV}\,=\,$80$\sqrt{10\,\mu G/B}$,
where $B$ is the postshock magnetic field strength.  If we assume that
$B\,=\,4\,B_{0}\,=\,$12$\mu$G, then we find that our estimated maximum energy and roll-off frequency lie well within
the limit allowed by the estimate of Hendrick \& Reynolds (2001).  

In a study of 25 LMC supernova remnants by
Mathewson \etal\, (1983), \s2 was found to exhibit the lowest radio flux density in
the sample ($<$ 0.02 Jy at 408 MHz).  This suggests that although \s2 accelerates cosmic rays to TeV energies, the 
density of these particles is significantly lower than that found in the other LMC SNRs.  Therefore, it seems unlikely
that the blast wave structure of \s2 is significantly altered by nonlinear feedback from cosmic ray acceleration.
Further evidence against efficient cosmic ray acceleration in \s2 may be found in our optical and {\it Chandra} data.  The detailed
correlation between the optical and X-ray morphologies (HGRS03, RGH03) suggests that the blast wave emission is predominantly 
thermal in origin.  In addition, the shock models of Boulares \& Cox (1988) predict precursor temperatures $\sim$10$^{6}$~K for
non-radiative shocks dominated by cosmic ray pressure.   This is clearly ruled out by the H$\alpha$ narrow component
width measurements of SRL94 and the presence of Balmer line emission from behind the shock.  

\section{SUMMARY AND DISCUSSION}

We have conducted a thorough spectral imaging study of the supernova remnant \s2 in H$\alpha$.
Spectra extracted from the blast wave show the broad and narrow component 
emission lines characteristic of non-radiative shocks in partially neutral gas.  The H$\alpha$ broad
component width ranges from $\sim$ 450 \kms along the eastern edge
to values $\sim$ 1000 \kms along the faint western edge.  No broad component is detected
in spectra of the faint northern rim, indicating a very low density in these regions and probably
a broad component width in excess of 1000 \kms.  The range of shock velocities is 500 \kms along
the eastern rim, 700$-$800 \kms along the southeastern and southern rims and 800$-$1000 \kms along the
western and northwestern rims.  other parts of the blast wave.  The H$\alpha$ broad-to-narrow
ratios \bn range from around 0.4 to 0.8, falling significantly below the values predicted by our shock models.
The most likely explanation for the discrepancy is extra narrow component emission from a cosmic ray precursor,
although a fast neutral precursor may also contribute some emission.

If collisional excitation in a 40,000~K cosmic ray precursor produces narrow line emission,
the steep Balmer decrement of the precursor should affect the H$\beta$ broad-to-narrow ratio
far less than the H$\alpha$ broad-to-narrow ratio.  Therefore, a measurement of the H$\beta$ broad-to-narrow
ratio in the future would be highly desirable.  Among the other studied Balmer-dominated remnants,
Tycho also exhibits an anomalously low H$\alpha$ broad-to-narrow ratio (SKBW91, Ghavamian 1999,
Ghavamian \etal\, 2001).  In that case, where both H$\alpha$ and H$\beta$ broad-to-narrow ratios were
measured, the former ratios were found to lie 40\% below the lowest predicted values, while the latter ratios were 
found to agree with the lowest predicted values.  This is consistent with detection of superimposed precursor and shock
emission.  Note that the high inferred preshock neutral fraction for
Tycho's SNR (GRHB00) would make the precursor emission more detectable there than remnants like
SN 1006, where the inferred preshock neutral fraction is only $\sim\,$10\% (Ghavamian \etal\, 2002).  A high
preshock neutral fraction and large size for \s2 (over twice that of Tycho) would certainly elevate the
role of a precursor in understanding the blast wave emission.

Our ability to map the shock velocity along the entire rim of \s2 has enabled us to estimate the age and
explosion energy of this remnant entirely from the blast wave kinematics. 
Assuming the blast wave propagates into a radially uniform ISM, the derived age for
\s2 is 3500$-$5000 years, less than half the value calculated from previous observations (SKBW91).
The explosion kinetic energy is (2$-$6)\,$\times$\,10$^{50}$ ergs, less than half the canonical value
assumed for Type Ia SNe.  One possible explanation for the difference may be the loss of energy during an earlier
phase of evolution, when \s2 experienced efficient cosmic ray acceleration.  However, the lack of evidence for nonthermal 
emission in {\it Chandra} X-ray spectra of the blast wave (RGH03), the strong similarity between 
the optical and X-ray morphologies of the blast wave and the faintness of the radio emission
(Mills 1984) suggest that blast wave energy loss to cosmic ray acceleration is not significant in the
current epoch.

The presence of significantly neutral gas around the entire blast wave of \s2 places a strong
limit on the amount of photoionizing radiation emitted by the accreting white dwarf progenitor. 
SuperSoft X-ray Sources (SSS) are believed to be one class of these progenitors (van den Heuvel
\etal 1992, Rappaport, Di Stefano \& Smith 1994).  In their detailed 
calculations of ionization conditions surrounding SSS, Rappaport \etal\, (1994) found that H is 
entirely ionized inside a radius of 28 pc for a source with standard parameters
T$_{eff}$\,=\,4$\times$10$^{5}$~K, L\,=\,10$^{38}$ ergs s$^{-1}$ in a medium with n$\,=\,$1
cm$^{-3}$.  This is clearly larger than the current blast wave radius of \s2 (R\,$\leq\,$9.5 pc) and limits
the luminosity of an SSS progenitor to values $\leq$\,10$^{37}$ ergs s$^{-1}$.  Note that due to the low ambient
density, the recombination time per H atom is $\sim\,$10$^{5}$ yr, so that little neutral gas should re-form
within the ionized bubble created by an SSS progenitor. Moreover, it is worth
noting that SSS progenitors with L $\gtrsim$ 10$^{37}$ ergs sm$^{-1}$ also appeared to be ruled out
in younger Balmer-dominated remnants of Type Ia SNe, such as SN 1006 and Tycho's SNR.  This limit is
especially stringent in the latter remnant, where the blast wave encounters significantly neutral gas
only 2$-$3 pc from the center of the explosion.

There are at least 4 other SNRs with nearly complete Balmer-dominated rims,
namely RCW 86 (Long \& Blair 1990, Smith 1997) in our
galaxy and 0509$-$67.5, 0519$-$69.0 and 0548$-$70.4 (Tuohy \etal 1982) in the LMC.  Of these remnants,
0509$-$67.5 shows no evidence of a broad component (Tuohy \etal 1982, SKBW91),
probably due to its very high shock speed.  On the other hand, although spectra of RCW 86 have
revealed strong broad component emission (Long \& Blair 1990, Ghavamian \etal\, 2001), it
is nearly 40\arcmin\, in size (Smith 1997).  This is larger than the FOV of most Fabry-Perot
spectrometers, making a detailed FP study of this SNR more time consuming and challenging.
Only 0519$-$69.0 and 0548$-$70.4 appear to be small enough ($<$\,1\arcmin) with small enough
broad component widths (750-1300 \kms, SKBW91) to be amenable to the type of
study we have undertaken of \s2.  By applying the blast wave kinematic analysis described
in this paper to 0519$-$69.0 and 0548$-$70.4, we would be able to enlarge our sample
of SNRs with well determined ages and explosion energies.  Such a study would be invaluable
in testing evolutionary models of young SNRs and in the interpretation of existing data from
the infrared to the X-ray regimes.

 P. G. would like to thank John Raymond for helpful discussions on the shock physics presented
in this work.  The authors also thank the referee for helpful suggestions on improving the paper.
This work was partially supported by Chandra Grants GO0-1035X and GO1-2052X,
and NSF grant AST 9619510.  C. E. R. acknowledges support from a NASA Graduate Student Research 
Program Fellowship.  J. P. H. and T. B. W. would like to thank the staff of CTIO for their 
hospitality and support during the Fabry-Perot observations.

\clearpage

\begin{figure}
\plotone{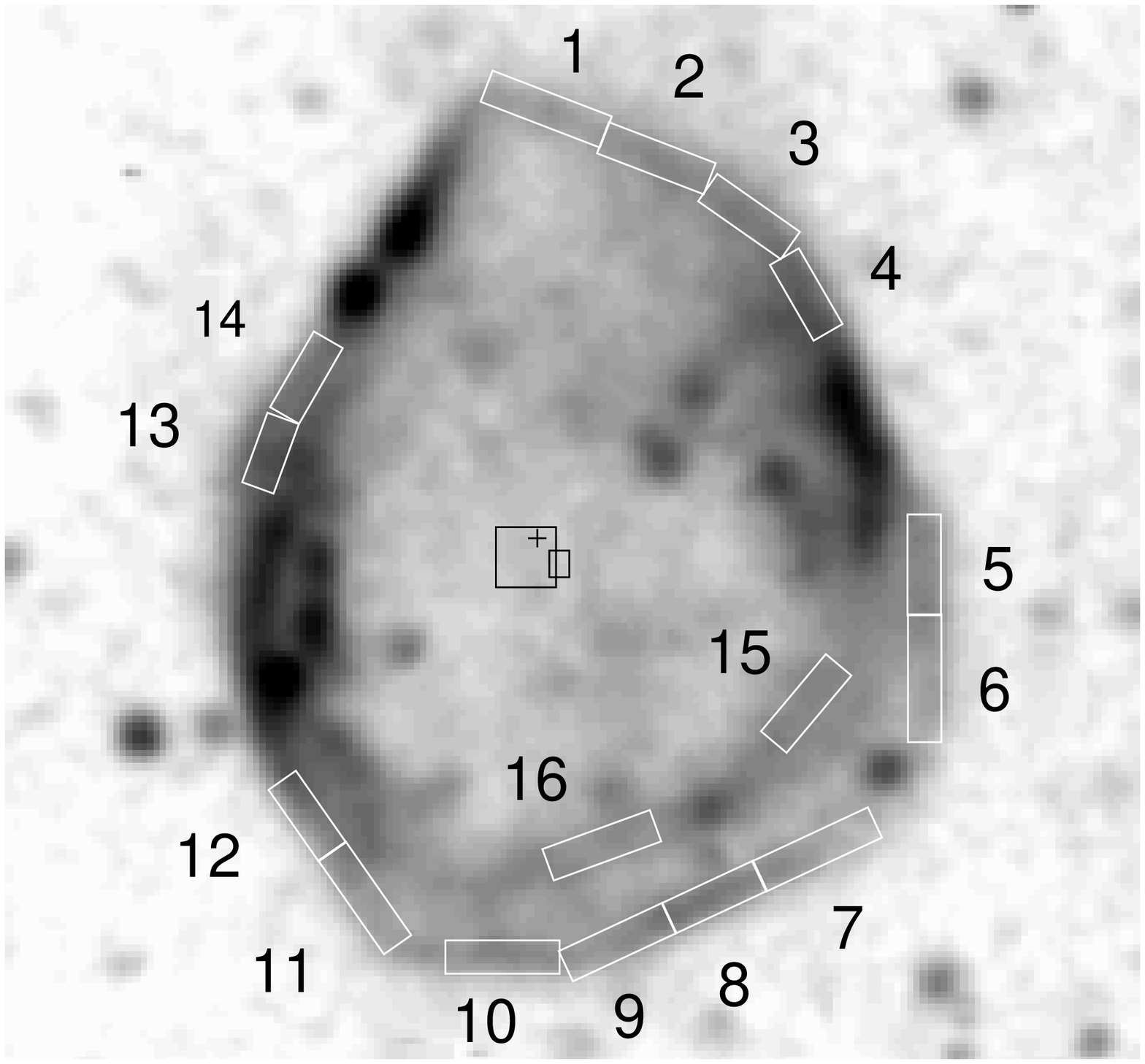}
\figcaption{The locations, sizes and orientations of spectral extraction apertures for the \s2\, blast wave.  The
image here is centered on the rest velocity of the narrow component H$\alpha$ line.  The width of each extraction
aperture is 3\farcs2.  The two black boxes near the center mark the 1$\sigma$
error bars on the derived geometric center from fits to $V_{S}(R)$ (larger box) and $n_{0}(R)$ (smaller box).  The
overlapping region between the two is 0\farcs7 $\times$ 2\farcs6 in size.  For comparison, the expansion center of
the iron-rich ejecta estimated from the {\it Chandra} data (HGRS03) is marked by the cross.  North is at the top, east is to the left. }
\end{figure}

\begin{figure}
\plotone{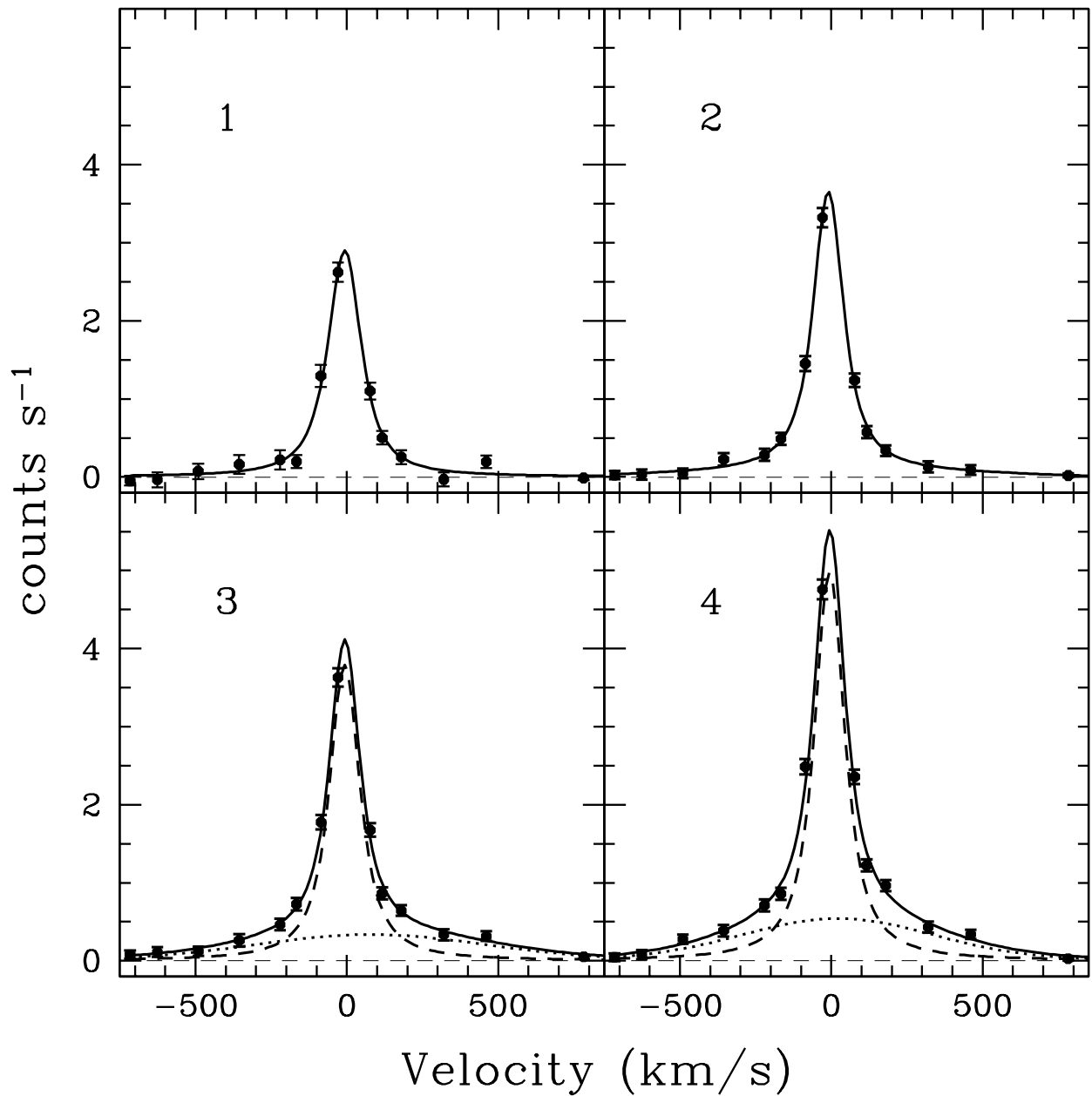}
\figcaption{The sky-subtracted H$\alpha$ line profiles of blast wave apertures 1 through 4.  The data are marked
by filled circles with error bars.  The total line profile fits are shown along with the individual broad and
narrow profiles.  The H$\alpha$ lines from apertures 1 and 2 are well fit by a single narrow component profile, suggesting
that the broad component is either nonexistent or too broad and faint to be detected. }
\end{figure}

\begin{figure}
\plotone{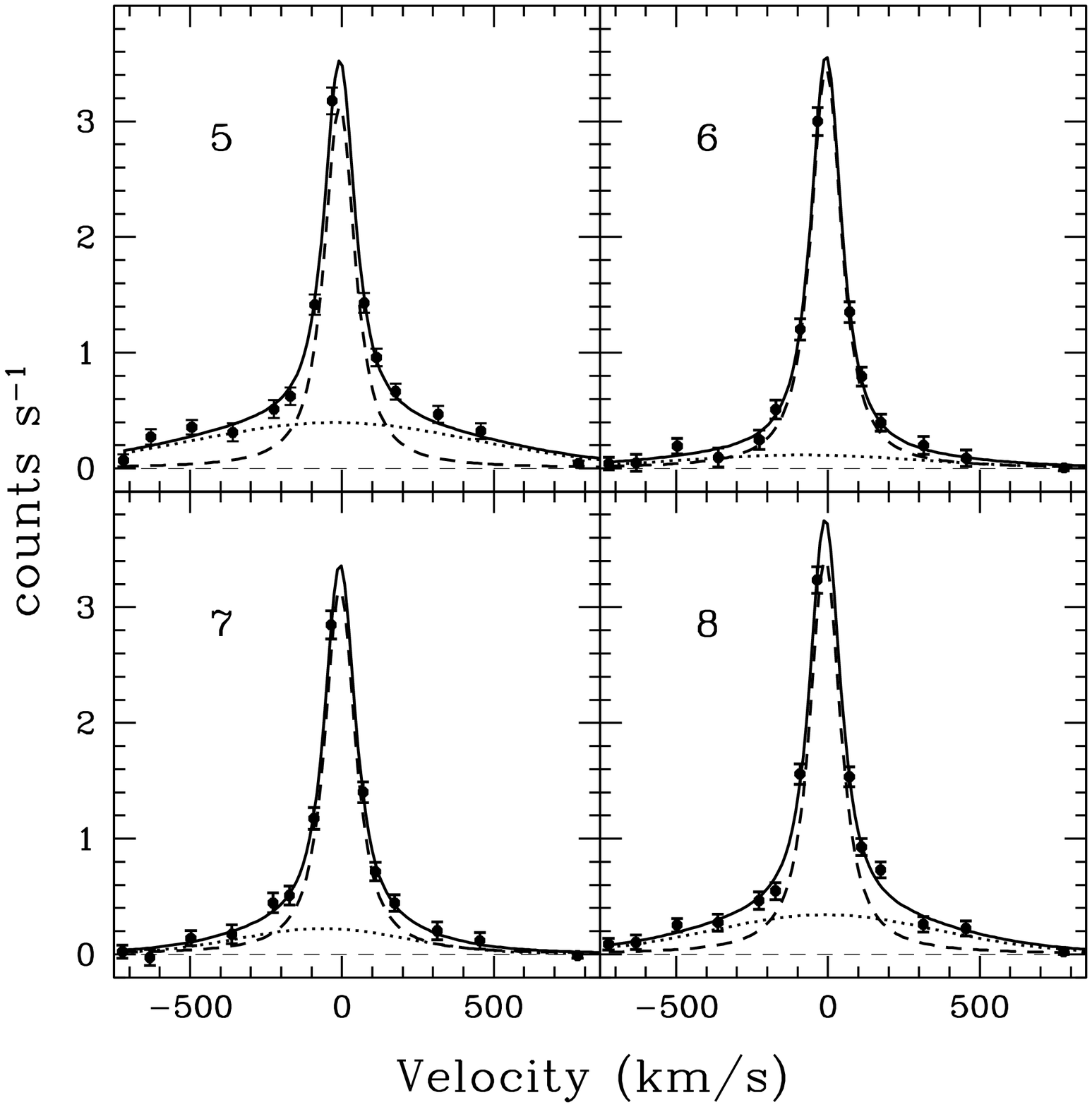}
\figcaption{The sky-subtracted H$\alpha$ line profiles of blast wave apertures 5 through 8.  
The total line profile fits are shown along with the individual broad and
narrow profiles.  }
\end{figure}

\begin{figure}
\plotone{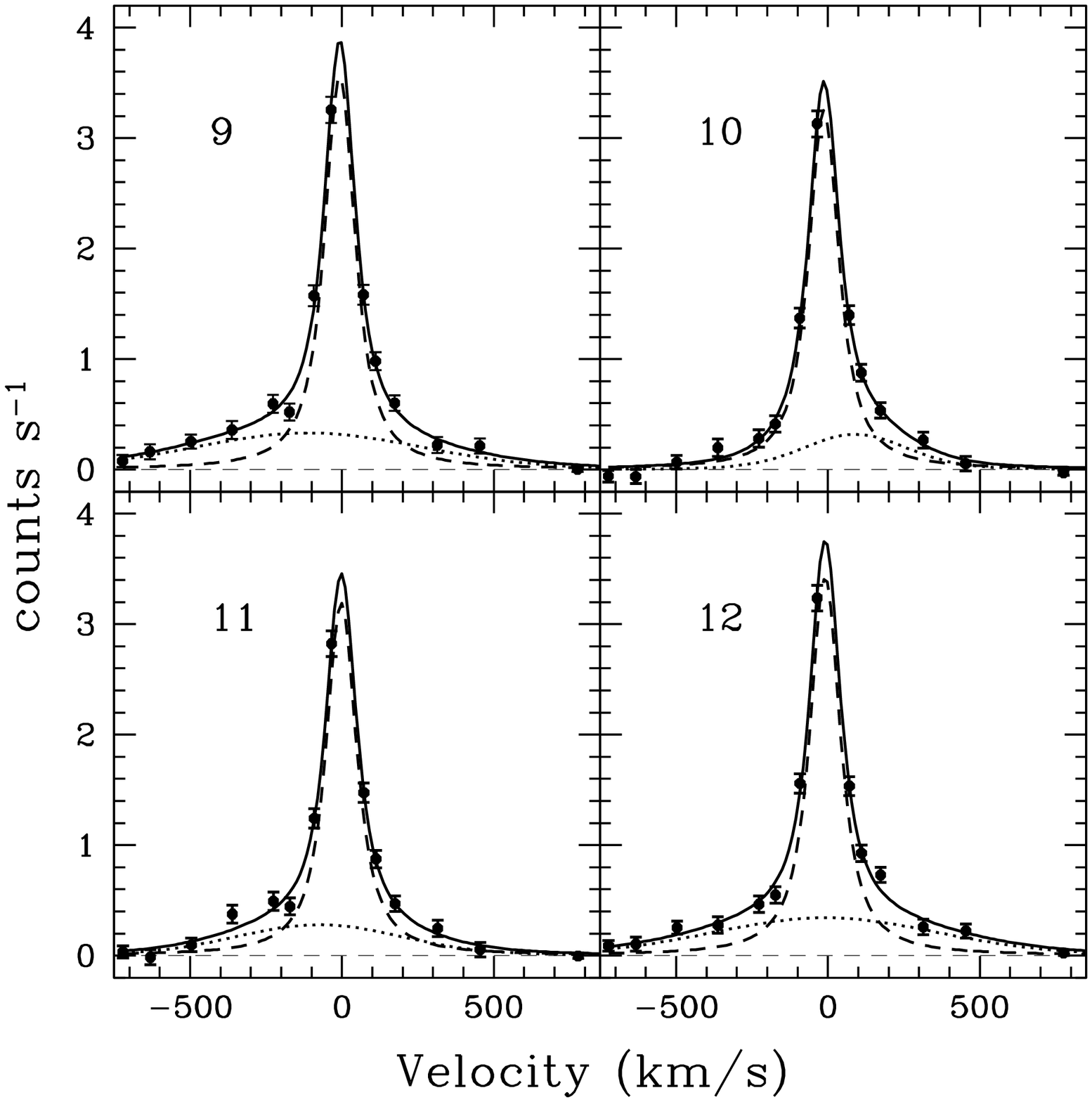}
\figcaption{The sky-subtracted H$\alpha$ line profiles of blast wave apertures 9 through 12. 
The total line profile fits are shown along with the individual broad and
narrow profiles.  }
\end{figure}

\begin{figure}
\plotone{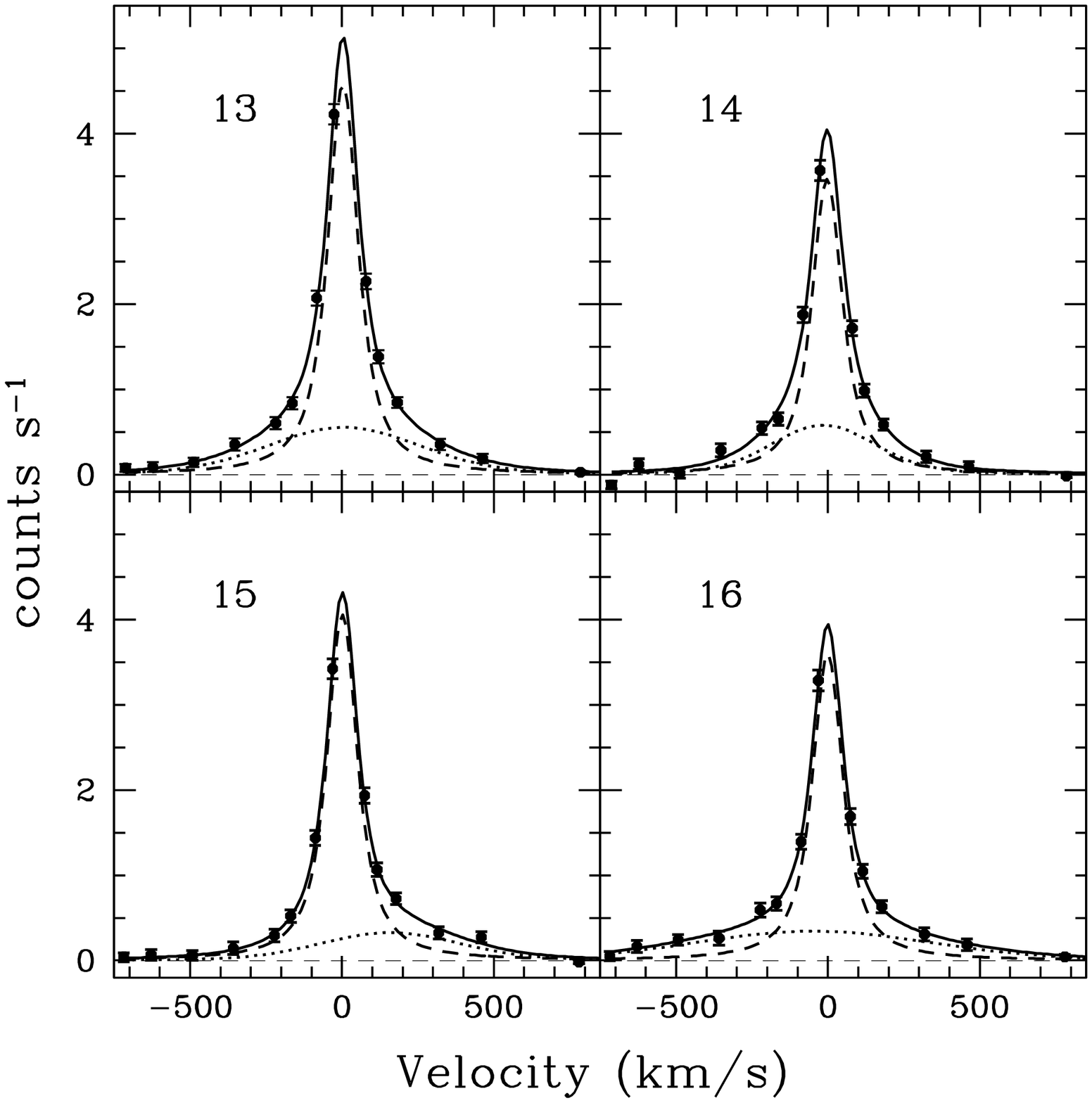}
\figcaption{The sky-subtracted H$\alpha$ line profiles of blast wave apertures 13 through 16.
The total line profile fits are shown along with the individual broad and
narrow profiles.  }
\end{figure}

\begin{figure}
\plotone{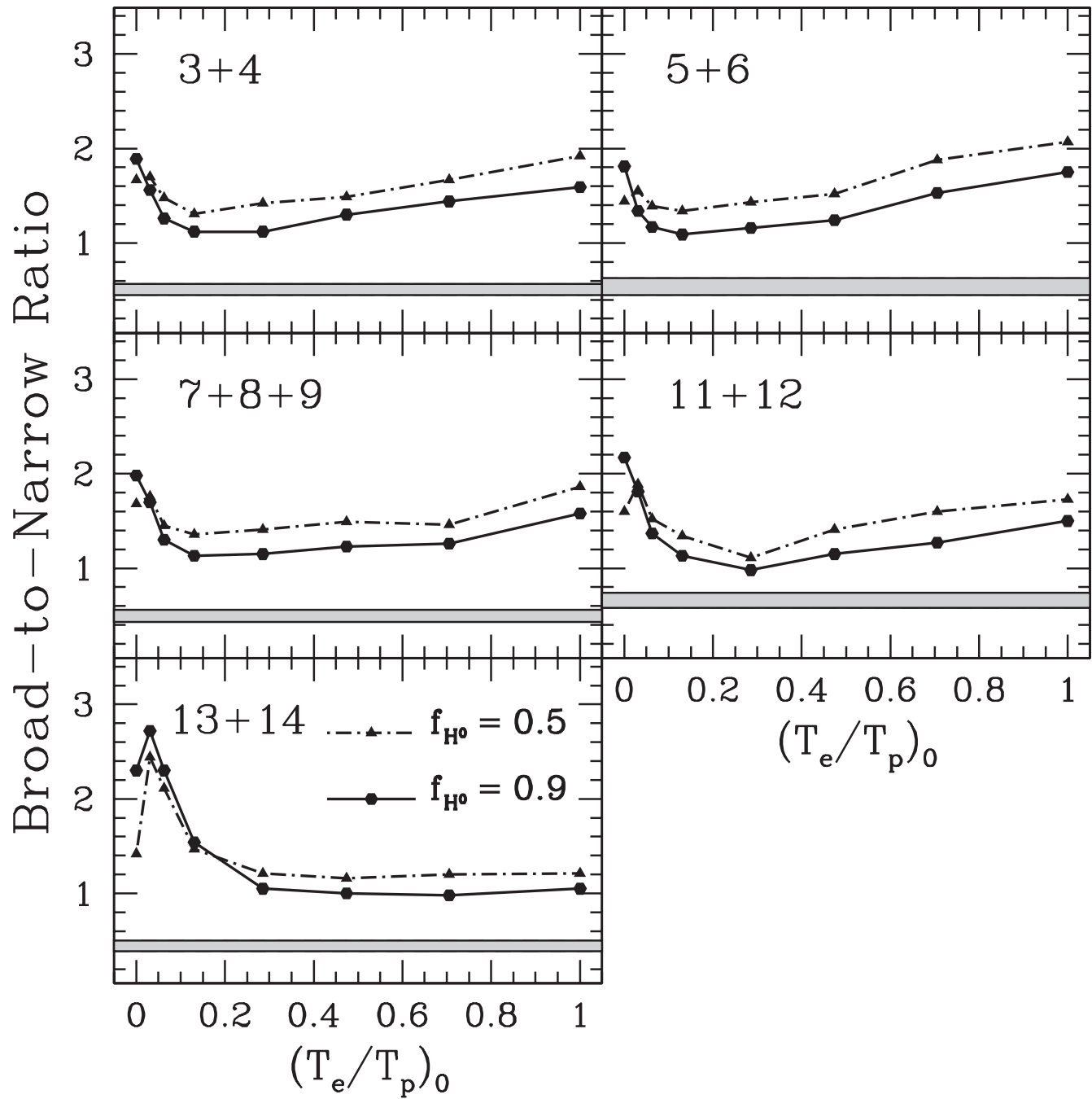}
\figcaption{The broad-to-narrow ratios predicted by shock models for combined blast
wave aperture regions.  The shaded regions indicate the observed range of H$\alpha$
broad-to-narrow ratios.  The curves are computed for a range of \tetp at the
shock front, for two selected preshock H neutral fractions: 50\% neutral H (dot-dashed curve) and 90\% 
neutral H (solid curve).  
}
\end{figure}

\begin{figure}
\plotone{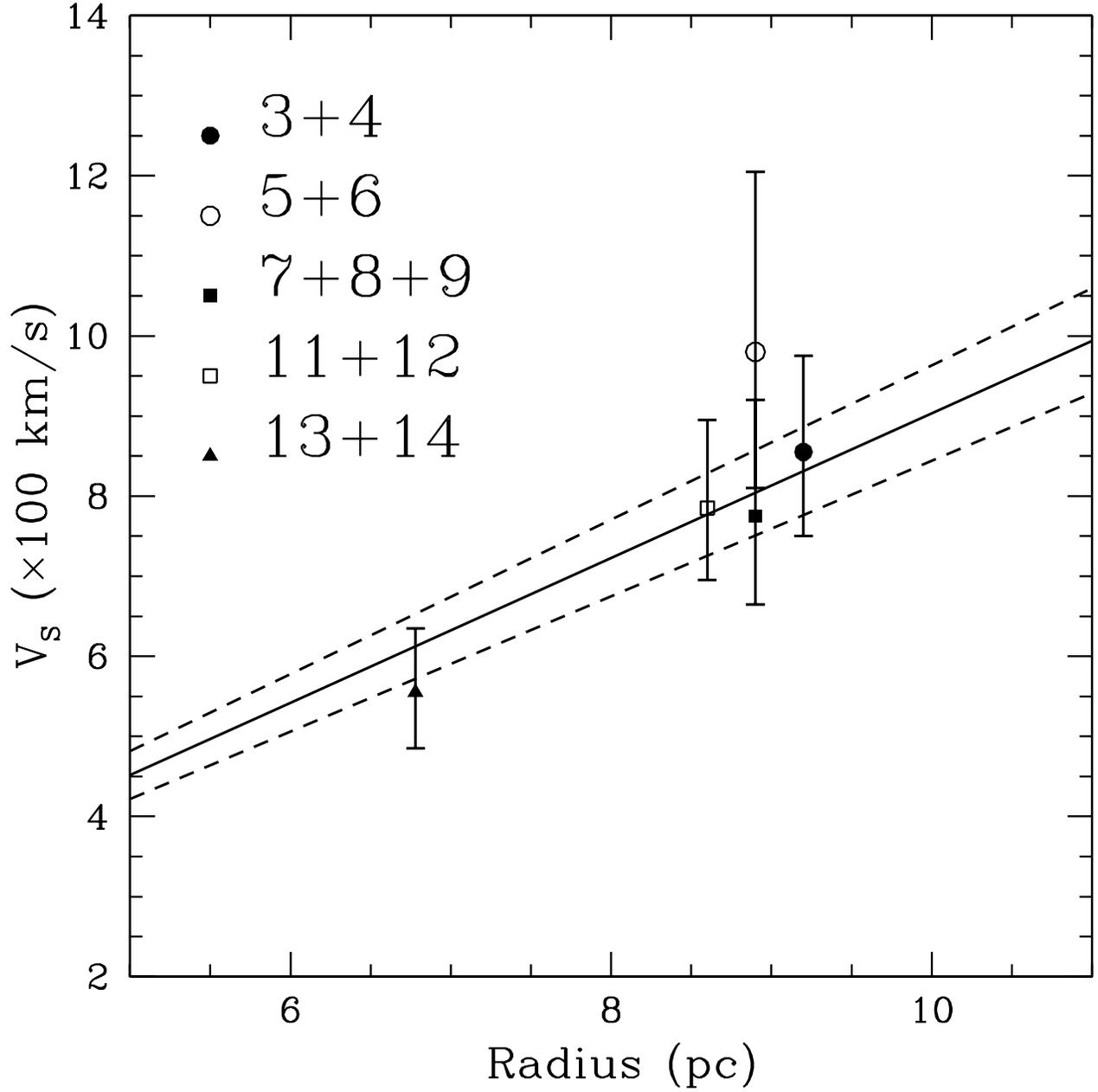}
\figcaption{Correlation between the blast wave speed on radius in \s2, shown for the combined apertures.
The best linear fit to the Sedov relation $V_{S}\,=\,\frac{2}{5}\,\frac{R}{\tau}$ is marked by the solid line.
The 1$\sigma$ limits on the slope are indicated by the dashed lines.
}
\end{figure}

\begin{figure}
\plotone{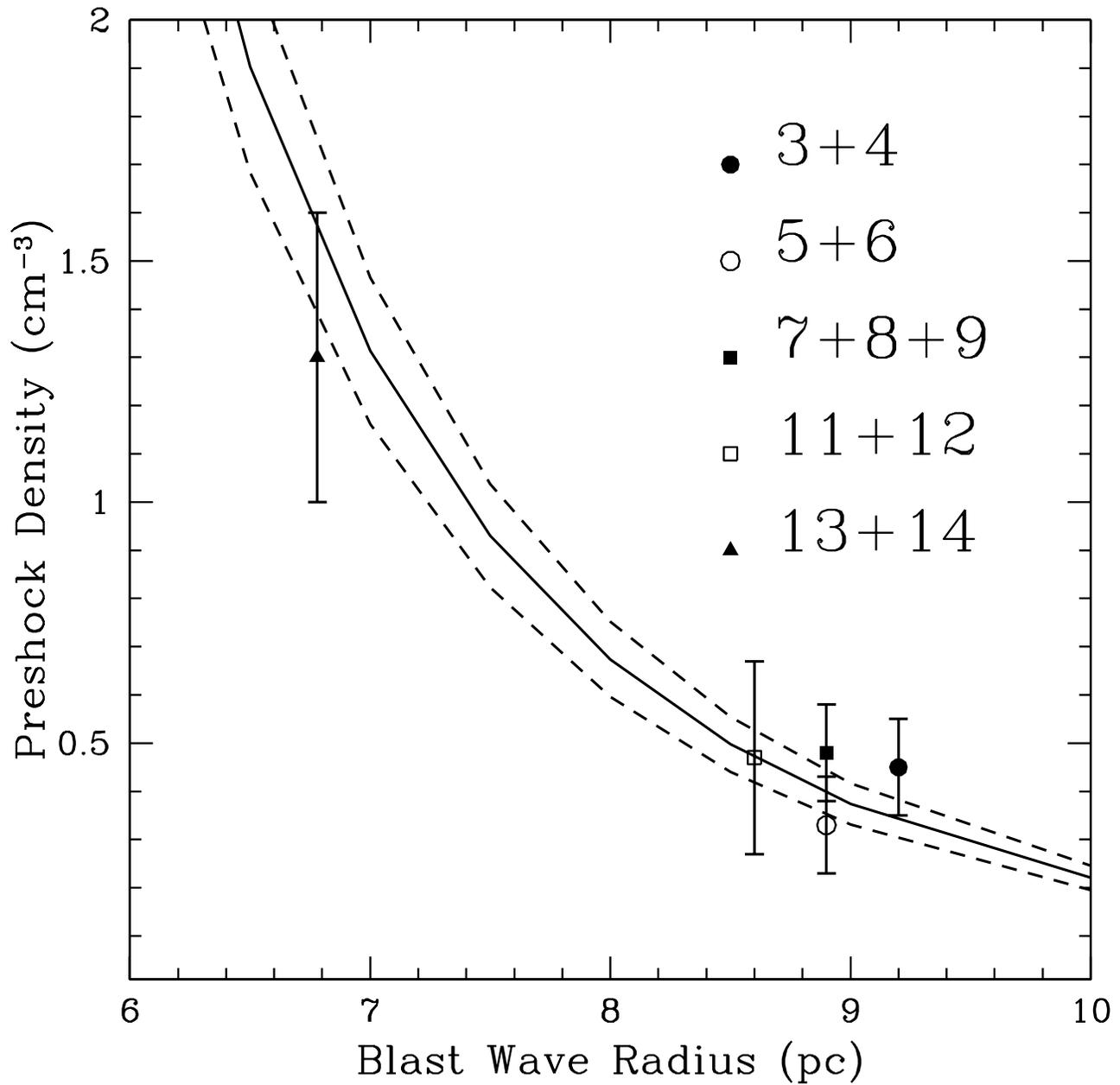}
\figcaption{
Correlation between the total preshock density and radius for \s2.   Preshock densities for the combined apertures
are determined from {\it Chandra} spectral models of RGH03.  The data are fit to the Sedov relation
$n_{0}(R)$ $\propto$ $R^{-5}$ (see Equation \ref{nvsr} in the text).  The solid line shows the best fit curve, 
while the dotted lines denote the 1$\sigma$ limits on the fit.
}
\end{figure}

\begin{deluxetable}{ccccccccc}
\tablenum{1}
\tablecaption{Blast Wave Extraction Apertures}
\tablehead{
\colhead{Aperture}   &
\colhead{$\alpha$(2000)} & \colhead{$\delta$(2000)} & \colhead{Size ($\arcsec$)} &
\colhead{Position Angle ($\circ$)}\tablenotemark{a}
}
\startdata
1  &   05:05:42.0  &  $-$67:51:57.5  &   3.2$\times$12.3  &  69\\
2  &   05:05:40.2  &  $-$67:52:02.6  &   3.2$\times$11.0  &  69\\
3  &   05:05:38.6  &  $-$67:52:08.5  &   3.2$\times$9.7  &  55\\
4  &   05:05:37.7  &  $-$67:52:16.2  &   3.2$\times$8.4	&  30\\
5  &   05:05:35.8  &  $-$67:52:42.8  &   3.2$\times$9.7	&  0\\
6  &   05:05:35.8  &  $-$67:52:53.8  &   3.2$\times$12.3 &  0\\
7  &   05:05:37.7  &  $-$67:53:10.1  &   3.2$\times$12.3 &  115\\
8  &   05:05:39.5  &  $-$67:53:14.4  &   3.2$\times$9.7 &  115\\
9  &   05:05:41.2  &  $-$67:53:18.5  &   3.2$\times$11.0 &  115\\
10 &   05:05:43.2  &  $-$67:53:19.7  &   3.2$\times$11.0 &  90\\
11 &   05:05:45.5  &  $-$67:53:13.5  &   3.2$\times$11.0  &  35\\
12 &   05:05:46.4  &  $-$67:53:05.4  &   3.2$\times$8.4 &  35\\
13 &   05:05:46.9  &  $-$67:52:30.1  &   3.2$\times$7.1 &  340\\
14 &   05:05:46.3  &  $-$67:52:22.9  &   3.2$\times$8.4 &  330\\
15 &   05:05:37.9 &   $-$67:52:55.9  &   3.2$\times$9.7  &  320\\
16 &   05:05:41.4 &   $-$67:53:09.1  &   3.2$\times$11.0  &  110\\
\enddata
\tablenotetext{a}{Measured E of N}
\end{deluxetable}

\begin{deluxetable}{ccccccccc}
\tablenum{2}
\tablecaption{Fitted Parameters for Blast Wave H$\alpha$ Profiles}
\tablehead{
\colhead{Aperture}   & \colhead{$\chi^{2}$/ 9 d.o.f.\tablenotemark{a}} &  \colhead{P($>$F)\tablenotemark{b}} &
\colhead{V$_{FWHM}$ (\kms)\tablenotemark{c}}  &
\colhead{I$_{B}$/I$_{N}$} &  \colhead{$\Delta V_{B-N}$\tablenotemark{d}}
}
\startdata
1	&	1.17	&	\nodata	&	\nodata	&	\nodata	&	\nodata\\
2	&	0.56	&	\nodata	&	\nodata	&	\nodata	&	\nodata\\
3	&	0.92	&	0.05	&	855$^{+165}_{-125}$	&	0.50$\pm$0.08	&  +85$^{+60}_{-51}$\\
4	&	2.59	&	0.05	&	700$^{+85}_{-75}$	&	0.51$\pm$0.06	&	+27$^{+29}_{-32}$\\
5	&	1.10	&	0.04	&	1055$^{+150}_{-120}$	&	0.88$^{+0.11}_{-0.10}$	&  $-$16$^{+51}_{-58}$\\
6	&	0.48	&	0.10	&	900$^{+775}_{-345}$	&	0.20$\pm$0.08	&  $-$80$^{+168}_{-335}$\\
7	&	0.54	&	0.07	&	595$^{+190}_{-150}$	&	0.29$\pm$0.09	& $-$59$^{+76}_{-88}$\\
8	&	1.26	&	0.05	&	830$^{+150}_{-125}$	&	0.55$^{+0.09}_{-0.08}$	& +2$^{+50}_{-54}$\\
9	&	1.09	&	0.05	&	835$^{+155}_{-120}$	&	0.51$^{+0.09}_{-0.08}$	& $-$103$^{+62}_{-52}$\\
10	&	1.16	&	0.11	&	350$^{+120}_{-95}$	&	0.26$\pm$0.09	& +104$^{+73}_{-52}$\\
11	&	1.07	&	0.07	&	595$^{+125}_{-100}$	&	0.36$\pm$0.09	&	$-$74$^{+60}_{-66}$\\
12	&	1.17	&	0.04	&	785$^{+95}_{-80}$	&	0.93$^{+0.11}_{-0.10}$	&  $-$74$^{+29}_{-37}$\\
13	&	0.60	&	0.04	&	540$^{+90}_{-80}$	&	0.46$^{+0.07}_{-0.06}$	&  +2$^{+24}_{-25}$\\
14	&	2.53	&	0.08	&	355$\pm$90		&	0.45$^{+0.16}_{-0.10}$	&  $-$13$^{+23}_{-28}$\\
15	&	0.81	&	0.06	&	490$^{+95}_{-85}$	&	0.28$\pm$0.07	&   +160$^{+63}_{-63}$\\
16	&	0.31	&	0.04	&	865$^{+180}_{-145}$	&	0.67$\pm$0.11	&	$-$48$^{+57}_{-62}$\\
\enddata
\tablenotetext{a}{The quoted $\chi^{2}$ values are for one-component profile fits for apertures 1 and 2 and
two-component profile fits for apertures 3-16.}
\tablenotetext{b}{P($>$F) denotes the F-test goodness-of-fit for inclusion of a broad component (see text for details).}
\tablenotetext{c}{The Gaussian FWHM of the broad component.}
\tablenotetext{d}{The shift of the broad component centroid from the narrow component centroid.}
\end{deluxetable}

\begin{deluxetable}{ccccccc}
\tablenum{3}
\tablecaption{H$\alpha$ Profile Fits For Combined Blast Wave Apertures}
\tablehead{
\colhead{Aperture}   & \colhead{$\chi^{2}$/9 d.o.f.} &  \colhead{P($>$F)\tablenotemark{a}} &
\colhead{V$_{\rm FWHM}$ (\kms)\tablenotemark{b}}  &
\colhead{I$_{\rm B}$/I$_{\rm N}$} 
}
\startdata
3+4	&	1.87	&	0.05	&	840$^{+115}_{-100}$	&	0.51$\pm$0.06\\
& & & & & & \\
5+6	&	0.68	&	0.04	&	985$^{+210}_{-165}$	&	0.54$\pm$0.09\\
& & & & & &  \\
7+8+9	&	0.73	&	0.04	&	805$^{+140}_{-115}$	&	0.49$^{+0.07}_{-0.06}$\\
& & & & & &  \\
11+12	&	0.40	&	0.04	&	735$^{+100}_{-85}$	&	0.66$\pm$0.08\\
& & & & &  & \\
13+14	&	1.90	&	0.04	&	450$\pm$60		&	0.44$^{+0.06}_{-0.05}$\\
\enddata
\tablenotetext{a}{P($>$F) denotes the F-test goodness-of-fit for inclusion of a broad component (see text for details).}
\tablenotetext{b}{The Gaussian FWHM of the broad component.}
\end{deluxetable}

\tablenum{4}
\begin{table}
\caption{Derived Shock Velocities for Combined Blast Wave Apertures}
 \begin{center}
    \begin{tabular}{cccccccccccc}\hline\hline
          \multicolumn{1}{c}{ }
        &\multicolumn{1}{c}{$(\frac{T_{e}}{T_{p}})_{0}$\,=\,$\frac{m_{e}}{m_{p}}$ } 
        & \multicolumn{1}{c}{$(\frac{T_{e}}{T_{p}})_{0}$\,=\,1} &
        \multicolumn{1}{c}{ } &  \multicolumn{1}{c}{ } & \multicolumn{1}{c}{ } &  \multicolumn{1}{c}{ {\it Chandra} Estimates\tablenotemark{a} } &
        \multicolumn{1}{c}{ }\\
        \hline
        \multicolumn{1}{c}{Aperture} 
        & \multicolumn{1}{c}{V$_{\rm S}$ (km s$^{-1}$)} &
        \multicolumn{1}{c}{V$_{\rm S}$ (km s$^{-1}$)} &  \multicolumn{1}{c}{ } & \multicolumn{1}{c}{ } & 
         \multicolumn{1}{c}{ V$_{\rm S}$ (km s$^{-1}$) } &
        \multicolumn{1}{c}{n$_{0}$ (cm$^{-3}$)} & \multicolumn{1}{c}{$(\frac{T_{e}}{T_{p}})_{0}$}\\
        \hline
& & & & &\\
3+4 (X1)\tablenotemark{b} & 815$^{\,+115}_{\,-100}$ & 1055$^{\,+140}_{\,-130}$  & & & 855$^{\,+120}_{\,-105}$  
& 0.45$^{+0.08}_{-0.12}$  &  0.24$^{+0.12}$ \\
& & & & &\\
5+6 (X2) & 960$^{\,+215}_{\,-165}$ & 1240$^{\,+290}_{\,-210}$  & & & 980$^{\,+225}_{\,-170}$  & 0.33$^{+0.05}_{-0.10}$  &  0.1$^{+0.29}$\\
& & & & &\\
7+8+9 (X3) &  775$^{\,+140}_{\,-115}$ & 1005$^{\,+170}_{\,-150}$  & & &  775$^{\,+145}_{\,-110}$  & 0.48$^{+0.05}_{-0.10}$  &  0.01$^{+0.03}$ \\
& & & & &\\
11+12 (X4) & 710$^{\,+100}_{\,-80}$  & 915$^{\,+130}_{\,-105}$   & & & 785$^{\,+110}_{\,-90}$  &  0.47$^{+0.09}_{-0.15}$  &  0.46$^{+0.18}_{-0.15}$ \\
& & & & &\\
13+14 (X5) & 430$^{\,+60}_{\,-55}$ & 555$^{\,+75}_{\,-70}$  & & & 555$^{\,+75}_{\,-70}$  & 1.30$^{+0.20}_{-0.26}$  &  1.0$_{-0.16}$\\
\end{tabular}
\end{center}
\tablenotetext{a}{\tetp is derived from X-ray spectral models of RGHW02. The appropriate shock speed for that
equilibration is quoted in column 4.}
\tablenotetext{b}{Names of the corresponding X-ray spectral extraction regions used in the \chan\,
analysis of RGHW02. }
\end{table}

\end{document}